\newcommand{\la}{\label}
\documentclass[preprint,singlecolumn,superscriptaddress,preprintnumbers,nopacs,floatfix,amsmath,amssymb]{revtex4}

\usepackage{graphicx}% Include figure files
\usepackage{dcolumn}% Align table columns on decimal point
\usepackage{bm}% bold math
\usepackage{amsmath}
\usepackage{graphicx}
\usepackage{amssymb}
\usepackage{mathrsfs}
\usepackage{bbm}
\usepackage{epsfig}
\newcommand{\be}{\begin{eqnarray}}
\newcommand{\ee}{\end{eqnarray}}

\begin{document}

%
%%%%%%%%%%%%%%%%%%%%%%%%%%%%%%%%%%%%%%%%%%%%%%%%%%%%%%%%%%
%
%\pagenumbering{empty}
%\begin{titlepage}
%
\title{ On Energy Functions for String-Like Continuous  Curves,  \\ Discrete Chains,  and Space-Filling One Dimensional Structures    }

%\vskip 5.0cm
\author{Shuangwei Hu}
\email{hushuangwei@gmail.com}
\affiliation{Department of Physics and Astronomy, Uppsala University,
P.O. Box 803, S-75108, Uppsala, Sweden}
\author{Ying Jiang}
\email{yjiang@shu.edu.cn}
\affiliation{Department of Physics, Shanghai University, 
Shangda Rd. 99, 200444 Shanghai, P.R. China}
\author{Antti J. Niemi}
\email{Antti.Niemi@physics.uu.se}
\affiliation{Department of Physics and Astronomy, Uppsala University,
P.O. Box 803, S-75108, Uppsala, Sweden}
\affiliation{
Laboratoire de Mathematiques et Physique Theorique
CNRS UMR 6083, F\'ed\'eration Denis Poisson, Universit\'e de Tours,
Parc de Grandmont, F37200, Tours, France}
\affiliation{Department of Physics, Beijing Institute of Technology, Haidian District, Beijing 100081, P. R. China}

\begin{abstract}
\noindent
The theory of string-like  continuous curves and discrete chains 
have numerous important physical applications.
Here we develop a general geometrical approach, to systematically derive  Hamiltonian 
energy functions for these objects. In the case
of continuous curves, we demand that the energy function must be invariant under local frame rotations,
and it should also transform covariantly under reparametrizations of the curve. This leads us
to consider energy functions that are constructed from the conserved quantities in the hierarchy of the integrable 
nonlinear Schr\"odinger equation (NLSE). We point out the existence of a
Weyl transformation that we utilize to introduce a  dual hierarchy to the standard NLSE hierarchy.
We propose that the dual hierarchy is also integrable, and we confirm this to the first non-trivial order.
In the discrete case the requirement of reparametrization invariance is void. But the demand of 
invariance under local frame rotations prevails, and we utilize it to introduce a discrete 
variant of the Zakharov-Shabat recursion relation.  We use this relation to derive frame independent
quantities that we propose are the essentially unique and as such natural
candidates for constructing energy functions for piecewise linear polygonal chains.  
We also investigate the discrete version of the Weyl duality transformation. 
We confirm that in the continuum limit the discrete energy functions go over to their 
continuum counterparts, including the
perfect derivative contributions. 
\end{abstract}

\maketitle

\section{Introduction}  

The theory of continuous curves in three dimensional space \cite{frenet}
is one of the pillars of differential geometry  \cite{spivak}. The study 
of string-like objects 
and their properties, both continuous and discrete, is similarly pivotal  to several apparently disparate subfields of physics.  
Examples include 
polymers \cite{polymer1}, \cite{polymer2}, Kirchoff-type elastic rods \cite{kirc1}-\cite{kirc3}, 
vortices in fluid dynamics, \cite{ricca1}, \cite{holm}, \cite{marsden}, turbulence \cite{turb},
superconductors \cite{nielsen1}, \cite{nielsen2}, superfluids \cite{volovik},  
plasmas \cite{ludvigp},  metallic hydrogen \cite{babaev},
and  confining \cite{conf1}, cosmic \cite{cosmic} and fundamental  \cite{fund} strings in high energy physics,
and numerous other applications. 
Strings can display intricate geometry including knots \cite{fadde}, strings can proceed by leapfrogging \cite{leap},
\cite{marsden},
and strings might even  realize exotic exchange statistics 
in three dimensions \cite{prl}.
Among the several  general level 
theoretical contributions,  we in addition
draw attention to
\cite{regge} and to \cite{poly} as particularly notable ones. 

In the present  article we shall be mainly  interested in deriving geometric
energy functions for discretized curves, or more precisely
piecewise linear polygonal chains in three space dimensions. 
The motivation comes from a recently proposed approach
to  the physics of proteins \cite{oma1}-\cite{oma5}. A protein is a  
biologically relevant example of a piecewise linear polygonal chain, 
with vertices identified as the central C$_\alpha$ carbons of the amino acids. 
A biologically active protein is also particularly interesting from the  point of view of physics of 
string-like structures in three space dimensions. It is an object that despite an inherent one dimensional 
character,  has physical properties  that causes it to collapse into 
a space filling construct. We propose that the study  of space filling, collapsed
string-like objects can have several important ramifications also in other areas of physics, from
understanding turbulent flows in fluid dynamics to potentially new forms of matter in high energy physics.

Our approach is based on the observation, first made by Hasimoto \cite{hasi1}, \cite{hasi2}, \cite{ricca1},
\cite{holm}  in the context of fluid dynamics, that the  one dimensional nonlinear Schr\"odinger 
equation (NLSE) describes string-like objects such as vortex filaments. For this he introduced a change of variables 
that relates the wave function of the NLSE to the Frenet frame representation of a space curve. 
The NLSE is a widely studied example of an integrable model \cite{fadtak}-\cite{appl} both due to its stature as a universal theoretical construct, and due to the
abundance of both theoretical \cite{fadtak}, \cite{abla} and practical  \cite{appl} applications. Among the
remarkable properties of the NLSE is the ability to support solitons as classical solutions. 
In the case of continuous curves, when the energy is computed by the canonical Hamiltonian of
the NLSE and the ensuing configuration is related to the geometry of a curve by the Hasimoto transformation, 
the soliton describes the buckling of the curve.

A properly  discretized version of the NLSE preserves the integrability  \cite{abla2}, \cite{kore}. It also supports 
soliton solutions that can be utilized to model the buckling of piecewise linear polygonal chains, 
via a discretized version of the Hasimoto transformation. This 
has been applied in \cite{cherno}, \cite{nora} 
to propose that the loop structures that are the characteristic geometric features of  
folded proteins, could be modeled using a properly
discretized version of the NLSE energy function.  In the case of proteins, the loops are 
solitons that cause the string-like C$_\alpha$ backbone to buckle
and collapse into a space filling configuration. As a consequence, from a general 
point of view, chains that support solitons with geometric interpretation
are truly curiosity arousing:  
Despite its inherently one dimensional character, a discrete string-like chain that supports
solitons athat cause it to  buckle,  appears like a localized space filling, particle-like extended object.

In the following,  we shall first outline how energy functions of continuous curves are derived from purely geometric
considerations, by demanding both invariance under local frame rotations and covariance under curve reparametrizations.
We show how  the NLSE together with its integrable hierarchy emerges naturally, when 
we demand that the energy function of a curve can not depend on the choice of framing along the curve. 
We scrutinize the properties of the Zakharov-Shabat \cite{zakh} recursion relation between the conserved quantities
in the NLSE hierarchy, from the point of view of frame independence. 
We observe that there is the possibility to introduce a dual hierarchy, which is related to the
original NLSE hierarchy by a Weyl scaling transformation. It turns out that 
the dual hierarchy has some quite attractive features. In particular,
it  embraces two of the conserved quantities of the NLSE hierarchy, the length and the helicity (Chern-Simons), 
that have no place in the standard NLSE 
recursion relations. We confirm by explicit computation that the leading order terms of the dual hierarchy are
mutually in involution, suggesting that the dual hierarchy gives rise to a novel  integrable set of conserved quantities.
But at the moment we still lack a general proof that the dual hierarchy is fully integrable {\it i.e.}
that all the infinitely many quantities that form this hierarchy are indeed in involution. 

We then proceed to the discrete case, of potentially space-filling chains. 
We first review the discrete generalization of the Frenet equations \cite{dff}.
We show by an explicit computation, that the two known integrable discretizations introduced in \cite{abla2} and \cite{kore}
coincide \cite{hoff}, by devising an explicit change of variables that maps these models to each other. We then proceed to
a discretization of the Zakharov-Shabat recursion relation. For this we adopt a geometric principle, that the discrete
version of the recursion relation should be formulated by demanding invariance under local frame rotation.
We introduce the ensuing discretizations of the conserved quantities, 
both in the case of the standard NLSE hierarchy and in the case of
its Weyl dual hierarchy. We confirm that in the continuum limit where the distance between the vertices in the discrete case
vanishes, the discretized versions of the conserved quantities indeed 
go smoothly over to their continuum versions,  including the perfect derivative contributions
which are necessary for the consistency of the higher order Zakharov-Shabat recursion relations.
We conclude with a discussion on the discrete energy functions that can be  utilized for example in numerical
studies, to describe the physical properties of piecewise linear polygonal chains, their
buckling,  and the ensuing space-filling  capacity.

%%%%%%%%%%%%%%%%%%%%%%%%%%%%%%%%%%%%%%%%%%%%%%%%%%%%%%%%%%
%
%
%
%
%
%
%
%
%
%
%
%
%
%
%
%%%%%%%%%%%%%%%%%%%%%%%%%%%%%%%%%%%%%%%%%%%%%%%%%%%%%%%%%%

\section{Framing And Reparametrizing Continuous Curves}

We consider a general, filamental, space curve $\gamma$ in $\mathbb R^3$ that can be described by a class $C^3$ differentiable
map $\mathbf x(z)$. Here $z \in [0,L]$ is  a generic parametrization and $L$ is the length of the curve,
\begin{equation}
H_{-1} \equiv L = \int\limits_0^L dz \,  \sqrt{ {\mathbf x}_z \cdot {\mathbf x}_z } \ 
\equiv \ \int\limits_0^L dz \, || \frac{d{\mathbf x}}{dz} || 
\la{L}
\end{equation} 
We note that this is essentially the standard time independent Nambu-Goto action of a string.
 For simplicity we shall assume that there are no inflection points along $\gamma$, so that
$ ||{\mathbf x}_z ||  \not= 0$. But  a generalization to include inflection 
points is straightforward \cite{dff}.

In the absence of inflection points, the curve $\gamma$ can be globally framed as follows. 
The unit length tangent vector is
\begin{equation}
\mathbf t \ = \   \frac{1}{ ||  {\mathbf x}_z|| } \,  {\mathbf x}_z 
\la{curve}
\end{equation}
It is orthogonal to the  unit length bi-normal vector
\[
\mathbf b \ = \ \frac{  {\mathbf x}_z\times  {\mathbf x}_{zz} } { || 
 {\mathbf x}_z \times {\mathbf x}_{zz} || }
\]
The unit length normal vector is
\[
\mathbf n = \mathbf b \times \mathbf t
\] 
The three  
vectors $(\mathbf n, \mathbf b, \mathbf t)$ form the right-handed orthonormal Frenet frame,  
at each point of the curve $\gamma$. 

The Frenet equation relates the frames at different points along $\gamma$. Explicitely, the
Frenet equation is \cite{frenet}, \cite{spivak}
\begin{equation}
\frac{d}{dz}\!\left(
\begin{matrix} 
{\bf n} \\
{\bf b} \\
{\bf t} \end{matrix} \right) =  \sqrt{g}  \left( \begin{matrix}
0 & \tau & - \kappa  \\ -\tau & 0 & 0 \\ \kappa & 0 & 0 \end{matrix} \right) 
\left(
\begin{matrix} 
{\bf n} \\
{\bf b} \\
{\bf t} \end{matrix} \right) 
\la{DS1}
\end{equation}

Here
\begin{equation}
\sqrt{g}  =  \frac{ds}{dz}  =  \sqrt{ {\mathbf x}_z (z) \cdot   {\mathbf x}_z (z) }  \ = \ || \mathbf x_z|| \ = \  \sqrt{g_{zz}} 
\la{prole}
\end{equation}
Since $\gamma$ is one dimensional, the metric on $\gamma$ has only one component that we denote $g_{zz}$. This
determines  the local scale of curve parametrization vis-$\grave{\rm a}$-vis the metric in $\mathbb R^3$.
Further, 
\begin{equation}
\kappa(z) \ = \ \frac{ || {\mathbf x}_z \times {\mathbf x}_{zz} || } { ||  {\mathbf x}_z||^3 }
\la{kg}
\end{equation}
is the (Frenet) curvature of $\gamma$ on the osculating plane that is spanned by $\mathbf t$ and $\mathbf n$, and
\begin{equation}
\tau(z) \ = \ \frac{ ( {\mathbf x}_z \times  {\mathbf  x}_{zz} ) \cdot { {\mathbf x}_{zzz} }} { || {\mathbf x}_z \times  {\mathbf x}_{zz} ||^2 }
\la{tau}
\end{equation}
is the torsion of $\gamma$.

If the metric,  curvature and torsion  
are known we can construct the frames by solving (\ref{DS1}), 
and then proceed to construct the curve  by solving (\ref{curve}). The solution is unique, up to rigid translations and 
rotations of the curve. In this manner the 
Frenet equation proposes us to construct energy functions for
curves in $\mathbb R^3$, in terms of the curvature and the torsion and the way how the curve is 
parametrized.

In the following we shall denote by $s \in [0,L]$ the arc-length parameter, while $z$ denotes generic parametrization.
The arc-length parameter $s$  measures
the length along $\gamma$ in terms of the distance scale of the three dimensional ambient space. 
The change of variables from a generic parameter $z$ to the arc-length parameter $s$ is
\[
s(z) = \int\limits_0^z || {\mathbf x}_z (z') || dz'
\]
Accordingly, we consider the effects of infinitesimal local diffeomorphisms 
along $\gamma$, obtained by deforming $s$ as follows
\begin{equation}
s \to  z = s + \epsilon(s)
\la{infi}
\end{equation}
 Here $\epsilon(s)$ is an arbitrary infinitesimally small function  such that
\[
\epsilon(0) = \epsilon (L) = 0 = \epsilon_s (0)  = \epsilon_s(L)
\]
The Lie algebra of diffeomorphisms (\ref{infi}) of a line segment in $\mathbb R^1$ is the 
classical Virasoro (Witt) algebra.   This proposes us to define a  
function $f(s)$ on $\gamma$ to have a weight $h$ akin the conformal weight, 
if $f(s)$ transforms according to
\begin{equation}
\delta f (s) = - \left( \epsilon \frac{d}{ds} +  h\epsilon_s \, \right) \, f(s)
\la{fe}
\end{equation}
under the infinitesimal diffeomorphism (\ref{infi}).  If $f$ has weight $h_1$ and $g$ has weight $h_2$
then the product $fg$ has weight $h_1+h_2$. 
We also note the finite version of (\ref{fe}),
\[
f(s) \ \to \  \tilde f (z) \ = \ \left( \frac{d t}{d s} \right)^{-h} \!\!  f(s)
\]

Since the three dimensional geometric shape of the curve $\gamma$ in $ \mathbb R^3$ does not
depend on the way how it has been parametrized, the embedding 
$\mathbf x(z)$  transforms as a scalar {\it i.e.} it has weight $h=0$
under reparametrizations.
Similarly, the curvature (\ref{kg}) and torsion (\ref{tau}) are
scalars  under reparametrizations. Infinitesimally, 
\[
\delta \kappa (s) = - \epsilon(s) \frac{d\kappa}{ds} \ \equiv \ -\epsilon {\kappa}_s
\]
\[
\delta \tau (s) = - \epsilon(s) \frac{d\tau}{ds} \ \equiv \ -\epsilon \tau_s
\]
Generic functions can be composed by taking derivatives, multiplying and summing 
up functions that 
have definite weights. 

If $f(s)$ is a  function on $\gamma$ that 
has a definite weight, in general its derivative does not
have a definite weight. 
For the derivative to acquire a definite weight,
we need to extend it into a covariant derivative along $\gamma$.
Given the metric $g_{zz}$ along $\gamma$, we can
deduce the covariant derivative  by considering its action on a function 
$f$ with weight $h$. For this we 
demand that under the infinitesimal transformation (\ref{infi}) 
\[
\left( \, \frac{d}{dz} + \Gamma(z) \, \right) \tilde f(z) - \left( \, \frac{d}{ds} +  \Gamma(s) \, \right)f(s)
\]
\[
= - \left( \, 
\epsilon \frac{d}{ds} + (h+1)  \epsilon_s  \, \right)
\left( \, \frac{d}{ds} +  \Gamma(s) \, \right) f(s) 
\]
Here $\Gamma$ is the connection on $\gamma$.  Explicitely, we can choose
\begin{equation}
\Gamma(z) = h \, \partial_z \ln \sqrt{g} \ \equiv \ h \, \partial_z \ln || \mathbf x_z(z) ||
\la{gamma1}
\end{equation}
where we have used (\ref{prole}). More generally, we may also choose 
\begin{equation}
\Gamma(z) \ \to \ \Gamma(z) + \mathcal A(z) 
\la{gamma2}
\end{equation} 
Here $\mathcal A(z)$ is a quantity with weight $h=1$. 

A general frame is related to the Frenet frame by 
a local SO(2) rotation, that sends
\begin{equation}
\left( \begin{matrix} {\bf n} \\ {\bf b} \end{matrix} \right) \ \to \ \left( \begin{matrix} {{\bf e}_1} \\ {\bf e}_2 \end{matrix} \right) \
= \ \left( \begin{matrix} \cos \eta(z) & - \sin \eta(z) \\ \sin \eta(z) & \cos \eta(z) \end{matrix}\right)
\left( \begin{matrix} {\bf n} \\ {\bf b} \end{matrix} \right)
\la{newframe}
\end{equation}
The ensuing rotated version of the Frenet equation is
\begin{equation}
\frac{d}{dz} \left( \begin{matrix} {\bf e}_1 \\ {\bf e }_2 \\ {\bf t} \end{matrix}
\right) =
\sqrt{g}  \left( \begin{matrix} 0 & \tau - \eta_z & - \kappa \cos \eta \\ 
- \tau + \eta_z  & 0 & -\kappa \sin \eta \\
\kappa \cos \eta & \kappa \sin \eta  & 0 \end{matrix} \right)  
\left( \begin{matrix} {\bf e}_1  \\ {\bf e }_2 \\ {\bf t} \end{matrix}
\right) 
\la{contso2}
\end{equation} 
We can write this as
\begin{equation}
\frac{d}{dz} \left( \begin{matrix} {\bf e}_1 \\ {\bf e }_2 \\ {\bf t} \end{matrix}
\right) \ = \ \sqrt{g}  \left( \begin{matrix} 0 & \tau_r & - \kappa_g \\ 
- \tau_r  & 0 & - \kappa_n \\
\kappa_g & \kappa_n  & 0 \end{matrix} \right)  
\left( \begin{matrix} {\bf e}_1  \\ {\bf e }_2 \\ {\bf t} \end{matrix}
\right) 
\la{kgn}
\end{equation}

\noindent
Here $\kappa_g$ is  the geodesic curvature, and $\kappa_n$ is the normal curvature, 
and $\tau_r$ is the relative torsion of the curve on a two dimensional surface $\mathcal S \in \mathbb R^3$; 
we obtain the surface
by deforming the osculating plane, around the point of contact with the curve, and different values of $\eta$ can also be
interpreted in terms of different choices of the surface $\mathcal S$.
Notice that while $\kappa_g$ and $\kappa_n$ in (\ref{kgn}) both depend on the surface, that is they are both frame dependent, 
the modulus
\begin{equation}
\rho \ = \ \bar \psi \psi = \bar\kappa \kappa = \kappa_g^2  + \kappa_n^2 
\la{rho}
\end{equation}
is a frame independent  characteristic of the curve $\gamma$. In particular, $\rho$ is a scalar under reparametrizations.

We  introduce the combination
\begin{equation}
\psi (z)  \ = \ \kappa(z) \exp\left(i\int\limits_0^{z} \! \tau \sqrt{g}  \,d z' \  \right) 
%= \kappa e^{i \int\limits^s \tau  }
\la{psi}
\end{equation}
Alternatively, we may introduce the more symmetric
\begin{equation}
\psi (z)  \ = \ \kappa(z) \exp\left( \, \frac{i}{2} \int\limits_0^{z} \! \tau \sqrt{g}  \,d z' \  + \frac{i}{2} \int\limits_L^z \tau \sqrt{g}  \,d z' \  \! \right) 
\la{psisym}
\end{equation}
The decomposition (\ref{psi})  
is essentially the Hasimoto variable of fluid dynamics \cite{hasi1}, \cite{hasi2}.  It has weight $h=0$,
and it is invariant under the frame rotation (\ref{contso2}): When
\begin{equation}
\tau(s[z]) \ \to \ \tau(s[z]) - \frac{d \eta}{ds} \ \equiv \  \tau (z) - \frac{1}{\sqrt{g}} \frac{d\eta}{dz}
\la{gautau}
\end{equation}
and 
\begin{equation}
\kappa \ \to \ e^{i\eta} \kappa
\la{gaukappa}
\end{equation}
we get
\begin{equation}
\psi(z) \ \to \ \kappa\, e^{i\eta(z)} \cdot \exp \left( i \int^{z} \!\! \sqrt{g} \,\tau - i \eta(z)  \right) \ \equiv \psi(z)
\la{gaukap}
\end{equation}
We also note that there are the following two natural, frame independent  realizations of the quantity $\mathcal A$ 
with weight $h=1$ in (\ref{gamma2}),
\begin{equation}
\mathcal A \ \sim \ j_z(z) = \partial_z \ln \psi 
\la{gama}
\end{equation}
and
\begin{equation}
\mathcal A \ \sim \ \bar j_z(z) = \partial_z \ln \bar \psi
\la{gamb}
\end{equation}
They both appear in our construction.
%
%
%
%
%
%
%
%
%
%
%
%
%
%
%
%%%%%%%%%%%%%%%%%%%%%%%%%%%%%%%%%%%%%%%%%%%%%%%%%%%%%%%%%%
%
%
%
%
%
%
%
%
%
%
%
%
%
%
%
%%%%%%%%%%%%%%%%%%%%%%%%%%%%%%%%%%%%%%%%%%%%%%%%%%%%%%%%%

%%%%%%%%%%%%%%%%%%%%%%%%%%%%%%%%%%%%%%%%%%%%%%%%%%%%%%%%%%
%
%
%
%
%
%
%
%
%
%
%
%
%
%
%
%%%%%%%%%%%%%%%%%%%%%%%%%%%%%%%%%%%%%%%%%%%%%%%%%%%%%%%%%%

\section{NLSE hierarchy of curves}

We are interested in frame independent energy functions of curves, 
that have definite and identifiable transformation properties under curve reparametrizations. 
We first observe that there are several frame independent energy functions, that have
a natural geometric 
interpretation. 
The length $H_{-1} \equiv L$ in (\ref{L}) is an example. Additional familiar examples are 
the total torsion (total helicity)
\begin{equation}
H_{-2} = \int\limits_0^L \tau(s) ds
\la{T}
\end{equation}
and the total squared curvature 
\begin{equation}
H_1 =  \frac{1}{2} \int\limits_0^L \kappa^2(s) ds
 = \frac{1}{2} \int\limits_0^L |\mathbf t_s|^2 ds
\la{K}
\end{equation}
The latter determines the Worm Like Chain (Kratky-Porod) model that has been extensively applied  {\it e.g.}
to study elastic properties of DNA \cite{busta}; for applications of (\ref{K}) 
to random surfaces, see \cite{poly}.
The Bernoulli elastic curve is defined by the following linear combination of (\ref{K}) and (\ref{L}),
\[
H_1 + H_{-1}  =  \frac{1}{2} \int\limits_0^L \{ \kappa^2(s) + \lambda \} ds
\]
Here $\lambda$ is a Lagrange multiplier that enforces the length constraint.
Finally, we also mention the following combination of curvature and torsion
\begin{equation}
H_K =  \frac{1}{2} \int\limits_0^L \{ \alpha \, \kappa^2(s) + \beta \, \tau^2(s) \}ds
\la{kt}
\end{equation}
This defines the Kirchhoff elastic energy of a filament,  that has also been studied
extensively in the literature\cite{kirc1}-\cite{kirc3}.
Here we have assumed the arc-length parametrization; recall that both $\tau$ and $\kappa$ are scalars
under reparametrizations.

In the context of three dimensional  fluid dynamics, the length  (\ref{L}) is commonly employed as the 
Hamiltonian of a vortex filament \cite{kirc1}-\cite{turb}. The
dynamics of the filament is given by the localized induction approximation
\begin{equation}
\frac{d\mathbf x}{dt} = \mathbf x_s \times \mathbf x_{ss} = \kappa \mathbf b
\la{lia2}
\end{equation}
where the time evolution follows from the Rasetti-Regge bracket \cite{regge}, with symplectic one-form
\[
\int \!ds\, dt \ \mathbf x \cdot \mathbf x_s \times \mathbf x_t
\]

Similarly, the energy (\ref{K}) has  the form of 
standard Hamiltonian of Heisenberg spin chain, with
$\mathbf t(s)$ the spin vector.  The dynamics is given by
\begin{equation}
\frac{d\mathbf t}{dt} = \mathbf t \times \mathbf t_{ss}
\la{hc2}
\end{equation}
and now the Poisson bracket is
\begin{equation}
\{ t^a(s), t^b (s') \} = -\epsilon^{abc} t^c (s) \delta(s-s')
\la{pbhs}
\end{equation}
Hasimoto \cite{hasi1}, \cite{hasi2} utilized the variable (\ref{psi}) to establish an equivalence between 
(\ref{lia2}) and (\ref{hc2}). For this he showed that
both are equivalent to the equation of motion of the non-linear Schr\"odinger  (NLSE) 
Hamiltonian
\begin{equation}
H_3 = \int ds \, \{  {\bar \psi}_s \psi_s + \lambda (\bar \psi \psi)^2 \}
\la{nlse}
\end{equation}
when the Poisson bracket is
\begin{equation}
\{ \psi (s) , \bar\psi(s') \}_s = i \delta (s-s') 
\la{ppbrac}
\end{equation}
Here we have introduced a generic coupling parameters $\lambda$, it
can be scaled to $\lambda = 1$ by suitable redefinitions. 

In the following we adopt (\ref{ppbrac}), its representations in terms of the
Hasimoto decomposition,  and its discrete version as the standard bracket
that we denote with the subscript $s$.

The NLSE Hamiltonian system (\ref{nlse}), (\ref{ppbrac})  
can be elegantly presented in a form which is manifestly  
covariant under reparametrizations of the curve: When we introduce an arbitrary 
parametrization  $s \to z$, the NLSE Hamiltonian becomes
\begin{equation}
H_3 = \int \sqrt{g} dz \, \left\{  \, g^{zz}  {\bar \psi}_z \psi_z +  
\lambda  (\bar\psi \psi)^2 
\,  \right\}
\la{h3cov}
\end{equation}
where we have used the definition (\ref{prole}) of the induced metric on $\gamma$.
The Poisson bracket is similarly reparametrization covariant,
\begin{equation}
 \{ \psi(z) , \bar\psi (z') \}_s = \frac{i}{\sqrt{g}} \delta (z-z')
\la{pbcov1}
\end{equation}

The NLSE equation is known to be integrable on the finite
segment $[0,L]$, when appropriate boundary conditions are imposed.
In fact,  the equations (\ref{lia2}), (\ref{hc2}) 
constitute a bi-hamiltonian pair that determines the NLSE hierarchy. 
Consequently  there is an infinite set of conserved charges in involution with respect to the ensuing brackets,
\begin{equation}
\{ H_n , H_m \} = 0 \ \ \ \ \ \ n,m \in \mathbb Z^+
\la{Hmn}
\end{equation}
The charges are integrals of densities 
\[
H_n \ = \ \int\limits_0^L \!\sqrt{g} \, dz \, \mathcal H_n(z)
\]
that are functionals of $\psi, \, \bar\psi$ and their derivatives. 

In general, we can expect that the densities in an integrable hierarchy 
are determined only up to an additive derivative contribution,
\begin{equation}
\mathcal H_n \ \simeq \ \mathcal H_n + d\Lambda
\la{per}
\end{equation}
where $\Lambda(z)$ is some functional of $\psi$ and $\bar \psi$ and their derivatives, and
subject to an integrable boundary condition such as 
$\Lambda(0)=\Lambda(L) $. Recall that in one dimension, any function can be 
represented as a derivative of its own integral.
Consequently, we limit our attention in (\ref{per}) 
to such functionals $\Lambda(z)$ that are {\it local } in the fields $\psi(z)$ and $\bar \psi(z)$
and their derivatives. In that case, we call $d\Lambda(z)$ a {\it perfect derivative}. 
In an integrable hierarchy the density $\mathcal H_n(z)$ of a conserved charge 
is in general an equivalence class, where two representatives are equivalent to 
each other when they share the same Poisson bracket relations 
and deviate from each other only by a perfect derivative. 
This equivalence relation defines a structure akin de 
Rham cohomology. A  mutually consistent choice of the perfect derivative contributions in
a hierarchy of conserved charges in involution, can be viewed as a "choice of gauge". 

Our aim is to utilize the NLSE hierarchy to introduce 
frame independent energy functions, in
a manner which is manifestly reparametrization covariant.  
Occasionally, we find it convenient to introduce $j_z(z)$ in (\ref{gama}) 
and 
\[
\rho(z) = \bar \psi(z) \psi(z)
\]
as the canonical variables, in lieu of the Hasimoto variables $\psi(s)$ and $\bar\psi(s)$.  Note that
alternatively the complex conjugate $\bar j_z(z)$ in (\ref{gamb}) could be used instead of $j_z(z)$, and 
even a  linear combination of the two could be utilized.
In terms of the variables $j_z$ and $\rho$, the NLSE Hamiltonian has the following manifestly covariant representation
\begin{equation}
H_3 \ = \ \int \!  \sqrt{g} dz \, \{ \, g^{zz} j_z \rho_z   - \rho \, g^{zz} j_z j_z + \lambda \rho^2\, \}
\la{h3}
\end{equation}
and the standard bracket (\ref{pbcov1}) is the inverse symplectic two-form
\begin{equation}
\Omega^{-1}_{\, \, \, \, \, 3}(z,z') \ = \ \left( \begin{matrix} \{ j_z(z) , j_z(z') \}_s & \{ j_z(z) , \rho(z') \}_s \\
\{ \rho(z) , j_z(z') \}_s & \{ \rho(z), \rho(z') \}_s \end{matrix} \right) 
\ = \ 
i \left( \begin{matrix} 0 & \frac{1}{\sqrt{g}} \frac{d}{dz} \\ \frac{d}{dz} \frac{1}{\sqrt{g}} & 0 \end{matrix} \right) \delta(z-z')
\la{brac2}
\end{equation}
We also record the ensuing NLSE equations of motion, in  terms of  the arc-length parameter:
\begin{equation}
\frac{d }{dt} \, j = \{ H_3, j(s) \}_s  = i\frac{d}{ds} \left[  \,  \frac{d}{ds}  j
-j^2  - 2\lambda \rho\,
\right]
\la{eqnh3}
\end{equation}
and
\begin{equation}
\frac{d}{dt}\, \rho = \{ H_3 , \rho(s) \}_s = -  i \frac{d}{ds} \left[(
\frac{d}{ds} - 2j ) \rho  \right]
\la{eqnh3b}
\end{equation}
We remark that both (\ref{eqnh3}) and (\ref{eqnh3b}) are perfect derivatives. 
Here and in the sequel we denote
\[
j \equiv j_s
\]
the variable $j_z(z)$, in arc-length parametrization. 

Instead of  (\ref{lia2}) and (\ref{hc2}), it is often more convenient to base the bi-Hamiltonian structure on  
(\ref{h3}), (\ref{brac2}), and with the 
conserved momentum 
\begin{equation}
P \equiv H_2 = \int  ds \, j\, \rho \ = \ \int \sqrt{g} dz \, g^{zz} j_z \rho
\la{h2}
\end{equation}
\[
\{ H_3, H_2 \}_s = 0
\]
as the second Hamiltonian, together with $H_3$. 
In the arc-length parametrization, the symplectic structure corresponding
to $H_2$ is
\[
\Omega^{-1}_{\, \, \, \, \, 2}(s,s') \ = \ \left( \begin{matrix} \{ j(s) , j(s') \} & \{ j(s) , \rho(s') \} \\
\{ \rho(s) , j(s') \} & \{ \rho(s), \rho(s') \} \end{matrix} \right) 
\]
\begin{equation}
= \  i \left( \begin{array}{cc}
 2 \lambda  \frac{d}{ds}  & - \frac{d}{ds} ( \frac{d}{ds} j) \\ \\ 
 (\frac{d}{ds} - j) \frac{d}{ds} & - (\frac{d}{ds}  \rho + \rho \frac{d}{ds}) 
\end{array} \right)   \delta(s-s')
\la{brac3}
\end{equation}
It is straightforward to confirm that with (\ref{h2}), (\ref{brac3}) we
obtain 
(\ref{eqnh3}), (\ref{eqnh3b}) as the equations of motion. 
The infinite number of conserved densities $\mathcal  H_n(s)$, $n=1,... $ of  the NLSE hierarchy can be iteratively 
constructed as follows,
\begin{equation}
\left( \begin{array}{c} \frac{ \delta \mathcal H_{n+1} }{\delta j(s) } \\ \\
\frac{\delta \mathcal H_{n+1} }{\delta \rho(s)}  \end{array} \right) \ = \ 
\Omega_3 \Omega^{-1}_{\, \, \, \, \, 2} \left( \begin{array}{c} \frac{\delta \mathcal H_{n} }{\delta j(s) }\\ \\
 \frac{\delta \mathcal H_{n} }{\delta \rho(s)} \end{array} \right)
\la{bih}
\end{equation}
and the construction is conventionally started from the conserved charge (\ref{K})
\begin{equation}
H_1 = \int ds \mathcal H_1 \ = \ \frac{1}{2} \int ds \,  \bar\psi \psi  \ \equiv \frac{1}{2}  \int ds \,  \kappa^2
\la{K2}
\end{equation}

We note that the length (\ref{L}) and the total torsion (\ref{T}) are also conserved charges in the NLSE hierarchy, even though
they do not have a natural place in the standard hierarchy relations; the square of the torsion that appears in (\ref{kt}) is
not the density of a conserved charge in the NLSE hierarchy. As  a consequence, both  (\ref{L}) and (\ref{T}) are
commonly interpreted as {\it additional} conserved 
charges, with negative order $H_{-1}$ and $H_{-2}$ respectively (see {\it e.g.} \cite{kirc2}, \cite{ricca1}), and  
included {\it ad hoc}.  We may also choose
\[
H_0 = 0
\]

Finally, we point out the following relation between the NLSE hierarchy 
and the classical Virasoro (Witt) algebra of 
diffeomorphisms on the segment $[0,L]  \in \mathbb R^1$. 
If we define
\begin{equation}
\rho (s) = \frac{1}{L^2}\sum\limits_{-\infty}^{\infty} \mathcal L_n e^{ 2\pi i n \frac{s}{L}}
\la{ln}
\end{equation}
and 
\begin{equation}
j(s) = \frac{1}{L^2} \sum\limits_{-\infty}^\infty \mathcal J_n e^{ 2\pi i n \frac{s}{L}}
\la{jn}
\end{equation}
and substitute in (\ref{brac3}) the coefficients obey the Poisson brackets
\begin{eqnarray}
\{ \mathcal L_n , \mathcal L_m \}_s & = & 2\pi (n-m) \mathcal L_{n+m} + \frac{cL}{12} (n^3 - n) \delta_{n,-m} ~~~~~
\la{alg1}
\\
\{ \mathcal L_n , \mathcal J_m \}_s &  = & - 2\pi m \mathcal J_{n+m} - i \, 4\pi^2  n^2 \delta_{n,-m} \,  L
\la{alg2} \\
\{ \mathcal J_n, \mathcal J_m \}_s  & = & -4\pi \hat{\lambda } n \delta_{n,-m} \, L 
\la{alg3} \\
\{ L, \mathcal L_n \}_s & = & \{ L , \mathcal J_n \}_s \ = \ 0
\la{alg4}
\end{eqnarray}
where (\ref{alg1})  is the classical Virasoro (Witt) algebra and $\mathcal J_n$ extends it with a current algebra; 
the central charge $c$ is  classically absent but in general it is
non-vanishing, in the quantum theory. Note that we have here interpreted  the length (\ref{L})  as
an additional element of the algebra; it is after all a conserved charge in the NLSE hierarchy. 
We have also renormalized the coupling $\lambda$ in (\ref{nlse}) as follows,
\[
\lambda \ \to \lambda(L) \ = \ L \, \lambda \ \equiv \ \hat \lambda 
\]
Moreover, 
\[
H_1 \simeq \int ds \,  \bar\psi \psi \ \equiv  \ \int ds  \rho \ \propto \ \mathcal L_0
\]
where we use (\ref{ln}). 
Since $\Omega_3$ in (\ref{brac2}) is field independent 
and $\Omega_2$ can be presented in terms of the current algebra 
generators  (\ref{alg1})-(\ref{alg3}) we conclude that all the classically
conserved charges 
$H_n$ are polynomials of the classical Virasoro-current algebra generators 
$\mathcal L_n$ and $ \mathcal J_n$, and length $L$.

%%%%%%%%%%%%%%%%%%%%%%%%%%%%%%%%%%%%%%%%%%%%%%%%%%%%%%%%%%
%
%
%
%
%
%
%
%
%
%
%
%
%
%
%
%%%%%%%%%%%%%%%%%%%%%%%%%%%%%%%%%%%%%%%%%%%%%%%%%%%%%%%%%

%%%%%%%%%%%%%%%%%%%%%%%%%%%%%%%%%%%%%%%%%%%%%%%%%%%%%%%%%%
%
%
%
%
%
%
%
%
%
%
%
%
%
%
%
%%%%%%%%%%%%%%%%%%%%%%%%%%%%%%%%%%%%%%%%%%%%%%%%%%%%%%%%%%

\section{Reparametrization covariance of recursion relations}

Besides (\ref{bih}), 
the NLSE hierarchy of conserved densities $\mathcal H_n(s)$ with $\ n \in \mathbb Z^+$ 
can be constructed using the following 
recursive relation  \cite{zakh},  \cite{fadtak}
\begin{equation}
\omega_{n+1}(s)  = - i\,  \frac{d\omega_n}{ds} + \lambda \bar \psi  \sum\limits_{j=1}^{n-1} \omega_j \omega_{n-j} 
\la{recu1}
\end{equation}
We assume that the arc-length parametrization is used, in the case of curves in $\mathbb R^3$. 
We also note that alternatively, the recursion could be based on the complex conjugate of (\ref{recu1}). 
When the construction in (\ref{recu1})  is started with 
\begin{equation}
\begin{matrix} \omega_0(s) & = & 0 \\
\omega_1 (s) & = & \frac{1}{4}\,  \psi(s)
\end{matrix}
\la{omega01}
\end{equation}
we  reproduce the 
densities $\mathcal H_n(s)$ of all the higher order conserved charges as follows \cite{fadtak}, \cite{abla},
\begin{equation}
\mathcal H_n(s)  = \bar \psi\,  \omega_n \ \ \ \ \ \ n \in \mathbb Z^+
\la{Hnp}
\end{equation}

For a generic parametrization $s \to z$ of the curve,
we need to covariantize the recursive relation (\ref{recu1}),  with due care. 
For this we remind that the Hasomoto variable 
$\psi(z)$ is a scalar {\it i.e.} has weight $h=0$ under reparametrizations. 
Thus we obtain the reparametrization covariant expression of the charge $H_1$ in (\ref{K})
by interpreting the ensuing density 
$\omega_1$ as a scalar. In a generic parametrization,
\[
\omega_1 (z) = \frac{1}{4}\, \psi(z)
\]
\[
H_1 = \int \sqrt{g} dz \, \mathcal H_1(z) \ = \  \int \sqrt{g} dz \, \bar \psi \, \omega_1
\]
Next, we aim to compute $\omega_2(z)$ from the proper reparametrization covariant version of (\ref{recu1}).
Since $\omega_1$ is a scalar, we conclude that we may set
\[
\omega_2 = - i \frac{d}{dz} \omega_1
\]
so that $\omega_2$ has weight $h=1$. In (\ref{h2}) we have displayed the corresponding
conserve charge $H_2$, in a covariant form.

We proceed to $H_3$. Since $\omega_1$ has weight $h=1$, for a covariant charge we 
need to interpret the derivative term in (\ref{recu1})
accordingly, as the weight $h=1$  covariant derivative using the connection (\ref{gamma1}).
The ensuing covariant version of (\ref{recu1}) then reproduces  (\ref{h3cov}):
\[
\omega_3 \ \equiv \ \omega_{31} + \omega_{32} \ = \ 
-i (\frac{d}{dz} + \Gamma ) g^{zz} \omega_2 + \lambda \bar\psi \omega_1^2 
\]
\[
=
-\frac{1}{\sqrt{g}} \partial_z (\sqrt{g} g^{zz} \partial_z \omega_1 ) + \lambda \bar\psi \omega_1^2 
\]
\[
\Rightarrow H_3 = \int  \! \!\sqrt{g} dz \, \bar\psi \omega_3
= \frac{1}{4}\!  \int \! \!\sqrt{g} dz  \left\{   g^{zz}  {\bar \psi}_z \psi_z +  
\lambda  (\bar\psi \psi)^2  \right\}
\]
Note that in $\omega_3(z)$, the first term $\omega_{31}$
has  weight $h=2$ while the second term $\omega_{32}$ has weight $h=0$. Consequently,
when we proceed to $H_4$ and onwards, and compute the derivative of $\omega_3$ {\it etc.} 
in the recursive relation (\ref{recu1}),  the derivative 
operator must be replaced by the appropriate covariant derivative:  In the case of $H_4$ we need to introduce 
a covariant derivative with weight $h=2$ when the derivative 
is acting on the first term $\omega_{31}$, and 
a covariant derivative with $h=0$ when acting on the second term in $\omega_{32}$,
\[
-i \frac{d}{ds} \omega_3 \ \to \ -i ( \partial_z + 2 \Gamma) \omega_{31} - i \partial_z \omega_{32}
\]
In this manner, by successively interpreting the derivative in (\ref{recu1}) as the appropriate covariant derivative,
we generate the manifestly reparametrization covariant versions of the
densities $\omega_{n+1}(z)$, order by order. These densities are
linear combinations of terms with different weights, and the highest weight term has weight 
$h=n$.

%%%%%%%%%%%%%%%%%%%%%%%%%%%%%%%%%%%%%%%%%%%%%%%%%%%%%%%%%%
%
%
%
%
%
%
%
%
%
%
%
%
%
%
%
%%%%%%%%%%%%%%%%%%%%%%%%%%%%%%%%%%%%%%%%%%%%%%%%%%%%%%%%%

%%%%%%%%%%%%%%%%%%%%%%%%%%%%%%%%%%%%%%%%%%%%%%%%%%%%%%%%%%
%
%
%
%
%
%
%
%
%
%
%
%
%
%
%
%%%%%%%%%%%%%%%%%%%%%%%%%%%%%%%%%%%%%%%%%%%%%%%%%%%%%%%%%%

\section{Hierarchy with positive order}

We first remind  that the densities $\mathcal H_n(s)$ 
are not necessarily uniquely determined. In particular,
there may be the latitude  (\ref{per}) to add a perfect derivative to the density. 
In the case of an integrable model, this latitude
can be restricted by imposition of proper (integrable) 
boundary conditions at the end points of the
segment $[0,L]$.  
Here, our goal is to utilize the NLSE hierarchy to find
frame independent and reparametrization covariant energy functions for curves. 
For this we note, that the end points of our curves can move freely in $\mathbb R^3$. 
As a consequence the boundary conditions that we need to impose at the end 
points $z=0$ and $z=L$, are  to be open.  In particular, we may deform 
the energy density of the curve
by  an appropriate perfect derivative term, as long as this term 
does not interfere with frame independence
and reparametrization covariance.  

From the point of view of algebraic relations in integrable models the canonical pair of complex
variables ($\psi, \bar\psi$) is adequate.
But we are interested in describing curves. For this, more work is necessary.  We need, 
in addition of solving the integrable equations, to construct a solution to
the Frenet equation (\ref{DS1}), (\ref{kgn}), followed by an integration of the curve 
reconstruction equation (\ref{curve}). For this reason we proceed to investigate
the recursion relation (\ref{recu1}) explicitly,
in terms of the (Frenet frame) curvature and torsion. For clarity, in the sequel 
we use the arc-length parametrization, a reparametrization
covariantization is straightforward following Section IV, and will be performed
when conceptually desirable.

We first observe that the relation (\ref{Hnp}) between the density $\mathcal H_n$ 
 and the quantity $\omega_n$ admits the following internal symmetry: The density remains intact,
 when we send
\begin{equation}
\begin{matrix}
\omega_n (s) & \to & e^{i\alpha(s)}  \omega_n(s) \\ \\
\bar \psi(s) & \to & \bar\psi (s) e^{-i\alpha(s)} 
\end{matrix}
\la{intsym}
\end{equation}
where $\alpha(s)$ is some function. Note that $\alpha(s)$  can also
depend on the canonical 
variables. In particular, (\ref{intsym}) does not need to be
a canonical transformation. 
The transformation (\ref{intsym}) converts 
the recursion relation (\ref{recu1}) into
\begin{equation}
\omega_{n+1}(s)  =  - \left( \,  i\frac{d}{ds} + \frac{d\alpha}{ds} \right) 
\omega_n + \lambda \bar \psi  \sum\limits_{j=1}^{n-1} \omega_j \omega_{n-j} 
\la{recu1b}
\end{equation}
We recall the Hasimoto variable (\ref{psi}) in the arc-length parametrization 
\[
\psi(s) = \kappa(s) \exp\left( i \int\limits_0^s \tau \,ds' \right)
\]
with ($\kappa,\tau$) the Frenet curvature and torsion, and choose
\[
\alpha(s) = - \int\limits_0^s \tau(s)\, ds'
\]
This gives us
\begin{equation}
\omega_{n+1}(s)  = - \left( \,  i \frac{d}{ds} - \tau(s) \right) 
\omega_n + \lambda \bar\psi \sum\limits_{j=1}^{n-1} \omega_j \omega_{n-j} 
\la{recu1c}
\end{equation}
which is manifestly covariant under frame rotations. 
  
We record the following five terms, for later reference; in computing these, we employ  the Frenet frame {\it a.k.a.}
unitary gauge relation (\ref{recu1}),  in 
which the curvature $\kappa(s)$ is a real valued quantity.
\begin{equation}
\begin{array}{lcl}
\omega_0  \ & = & \ 0 \\ \\
\omega_1 & = & \ \frac{1}{4}\kappa \\ \\
\omega_2  \ &  = &\   \frac{1}{4}\, \tau \kappa - \frac{i}{4} \kappa^\prime \\ \\
\omega_3 \ & = & \ - \frac{1}{4} \kappa^{\prime \prime} + \frac{1}{4} \tau^2 \kappa + 
\frac{ \lambda}{16}  \kappa^3 - \frac{i}{4} \left(
\tau^\prime \kappa + 2 \tau \kappa^\prime \right) \\ \\
\omega_4   \  &  =  & \frac{1}{16} \left( \,4 \kappa \tau^3 - 4 \kappa \tau^{\prime \prime}
- 12 \tau \kappa^{\prime \prime} - 12 \tau^\prime \kappa^\prime +  3 \lambda \kappa^3 \tau \, \right) \\ \\
&& \!\!\! + \, \frac{i}{16} \left( \, 4 \kappa^{\prime\prime\prime} - 12 \tau \tau^\prime \kappa - 12 \kappa^\prime \tau^2
- 5 \lambda \kappa^2 \kappa^\prime \right) \\ \\
\end{array} 
\end{equation}
We multiply these with $ \bar \kappa $ to obtain the ensuing versions of the 
integrable densities $\mathcal  H_n$; we remind in the Frenet frames the curvature is real so that 
$\bar\kappa = \kappa$. We list the following four for later reference: 
\begin{equation}
\begin{array}{lcl}
\mathcal H_1 & = & \ \frac{1}{4}\kappa^2 \\ \\
\mathcal H_2  \ &  = &\   \frac{1}{4}\, \tau\kappa^2  - \frac{i}{8} \frac{d}{ds} (\kappa^2) \\ \\
\mathcal H_3 \ & = & \  \frac{1}{4}( \kappa^{\prime})^2 + \frac{1}{4} \tau^2 \kappa^2 + \frac{\lambda}{16} \kappa^4 -
\frac{i}{4} \frac{d}{ds}\left( \tau \kappa^2 - i \kappa \kappa^\prime \right)
\\ \\
\mathcal H_4   \  &  =  & \frac{i}{4} \kappa \kappa^{\prime\prime\prime} \\ \\
& & \!\!\! + \frac{1}{16} \left( \, 8 \kappa 
\kappa^{\prime \prime}\tau + 4 \kappa^2 \tau^3 - 4 (\kappa^\prime)^2 \tau + 3 \lambda \kappa^4 \tau \, \right)
\\ \\
& & \!\!\! + \, \frac{i}{64} \frac{d}{ds} \left( \,  -  24 \kappa^2 \tau^2 + 5 \lambda \kappa^4 + 16 i \left[ \,  \kappa^2 \tau^\prime  + 
\kappa \kappa^\prime\tau \, \right] \right)~~~ \\ \\
\end{array} 
\la{hier4}
\end{equation}
We identify in $\mathcal H_1$ the mass (number) density
of NLSE. In $\mathcal H_2$ we identify the canonical momentum density, in addition of a perfect derivative term.
In $\mathcal H_3$ we have the NLSE Hamiltonian density, in addition of a perfect derivative term. Finally, $\mathcal
H_4$ is the complex modified KdV density together with a perfect derivative term. Note that the form of the perfect derivative terms
is fully dictated by the recursion relation. Even though the presence of a perfect derivative does not contribute to
the charge at the given order, it does influence the functional form of higher order conserved charges.

We can also present the recursion relation directly in terms of the integrable densities. The result is 
\begin{equation}
\begin{array}{lcl}
\mathcal H_0  \!\! \!& = & \!0 \\ \\
\mathcal H_1 \!\! \! &  = &  \! \frac{1}{4} \kappa^2 \\ \\
\mathcal H_{n+1} \! \!\! &  =  & \!  - i \, \mathcal D \, \mathcal H_n + \lambda \sum_{j=1}^{n-1} \mathcal H_j \mathcal H_{n-j}  \\ \\
\mathcal D \!\!\!  & = & \! \frac{d}{ds} + \left(\,  i \tau -  [\ln {\kappa}]^\prime\,\right)   = \frac{d}{ds} - \frac{d}{ds} \ln \bar \psi 
\equiv \frac{d}{ds}- {\bar j}(s) 
\end{array}
\la{posu}
\end{equation}
It is notable that the $h=1$ quantity (\ref{gamb}) appears.

%%%%%%%%%%%%%%%%%%%%%%%%%%%%%%%%%%%%%%%%%%%%%%%%%%%%%%%%%%
%
%
%
%
%
%
%
%
%
%
%
%
%
%
%
%%%%%%%%%%%%%%%%%%%%%%%%%%%%%%%%%%%%%%%%%%%%%%%%%%%%%%%%%

%%%%%%%%%%%%%%%%%%%%%%%%%%%%%%%%%%%%%%%%%%%%%%%%%%%%%%%%%%
%
%
%
%
%
%
%
%
%
%
%
%
%
%
%
%%%%%%%%%%%%%%%%%%%%%%%%%%%%%%%%%%%%%%%%%%%%%%%%%%%%%%%%%%

\section{Hierarchy with negative order}

We observe that the derivative operator $\mathcal D$  that appears in (\ref{posu})
and its conjugate
\[
{\bar{\mathcal D}} = \frac{d}{ds} + j(s)
\]
are related by the Weyl transformation,
\begin{equation}
{\bar {\mathcal D}} \ = \  e^{-\theta} \mathcal D e^{\theta} 
\la{sca1}
\end{equation}
when we choose
\begin{equation}
e^\theta = |\psi|^2
\la{sca2}
\end{equation}
This symmetry between the two quantities (\ref{gama}) and (\ref{gamb}) is most natural,
and proposes us to introduce the following extension (Weyl dual) of (\ref{recu1}),
for all negative integers: We simply map
\[
\omega_n \ \longrightarrow \ \omega_{-n} = e^{-\theta} \omega_n
\] 
This gives
\[
\omega_{-(n+1)} = -i \left( \frac{d \omega_{-n}}{ds} 
+\omega_{-n} \frac{d \ln |\psi|^2}{ds} \right) 
+ \lambda \bar\psi |\psi|^2 
\sum\limits_{j=1}^{n-1} \omega_{-j} \omega_{-(n-j)}
\]
with 
\[
\omega_{-1} = \frac{1}{\bar\psi}
\]
We define 
\[
\mathcal H_{-n} = \bar\psi \,\omega_{-n}
\]
to arrive at the following recursion relations of densities
\begin{equation}
%\left\{ 
\begin{array}{lcl}
\mathcal H_0  \ & = & \ 0 \\ \\
\mathcal H_{-1}  \ &  = &\  1 \\ \\
\mathcal H_{-(n+1)}   \  &  =  & \  - i {\bar {\mathcal D}} \, 
\mathcal H_{-n} + \lambda |\psi|^2 \sum_{j=1}^{n-1} \mathcal H_{-j }\mathcal
H_{-(n-j)} \\ \\
{\bar {\mathcal D}} \ & = & \ \frac{d}{ds} +  j(s)  \ \equiv \ \frac{d}{ds} + \frac{d}{ds} \ln \psi 
\end{array}
%\right.
\la{nesu}
\end{equation}
%
%
%\begin{equation}
% \begin{array}{lcl}
%\mathcal H_{-n}  \  &  =  & \ e^{-\theta} \mathcal H_{n}  \\
%{\bar {\mathcal D}} \ & = & \  e^{-\theta} \mathcal D e^{\theta} \\
%e^\theta \ & = & \  |\psi|^2  
%\end{array}
%\la{nesu2}
%\end{equation}
where 
\begin{equation}
\mathcal H_{-n} = e^{-\theta}\mathcal H_n
\la{sca3}
\end{equation}
We present the   following three negative order densities, for reference and scrutiny, using the Frenet frame:
\begin{equation}
\begin{array}{lcl}
\mathcal H_{-2} (s)  &  =  &  - i \frac{d}{ds} \ln \psi    =  \tau - i \frac{d}{ds} \ln \kappa
\\ \\
\mathcal H_{-3} (s) \ & = & \   -  \frac{d^2}{ds^2}   \ln \psi - (\frac{d}{ds} \ln \psi)^2 + \lambda |\psi|^2   \\ \\
& = & \left[ \tau - i (\ln \kappa)^\prime \right]^2 +1 + \lambda \kappa^2 -  i \left[  \tau + i (\ln \kappa)^\prime \right]^\prime \\ \\
\mathcal H_{-4}(s)  \ & = & \  \frac{i}{\psi} \frac{d^3}{ds^2} \psi - i \lambda ( \psi \frac{d}{ds} \bar\psi + 4 \bar\psi \frac{d}{ds} \psi)
\end{array}
\la{hesu}
\end{equation}
Remarkably, 
the two conserved charges (\ref{L}) and (\ref{T}) of the NLSE hierarchy now appear naturally, as quantities in the
negative order recursive relation. We remind that in a
conventional approach to NLSE, 
these two charges do not arise from the recursion relations (\ref{recu1}). Instead, they
are  introduced in an {\it ad hoc} manner \cite{kirc2}, \cite{ricca1} as additional charges.
Furthermore,
the density $\mathcal H_{-3}$ can be given
a familiar interpretation in terms of quantities that relate to gauge theories. This density is like 
the gauge invariant Proca mass for $\tau$,  when we interpreted $\tau$ as a gauge field and
$\ln \kappa$ as the ensuing Stueckelberg field;
the combination is gauge invariant under the analog U(1) gauge transformation  (\ref{gautau}), (\ref{gaukappa}).
See also next Section. 

Unlike (\ref{L}) and (\ref{T}), 
the integral of $\mathcal H_{-3}$ is not a conserved quantity in the NLSE hierarchy. For example, we find, using the
standard bracket (\ref{ppbrac})
\[
\{ H_3 , H_{-3} \}_s =2 i \int \! ds \left[ \, \frac{( \ln \psi)^{\prime\prime} 
(\psi)^{\prime\prime}}{\psi} - 2 \lambda |\psi|^2 (\ln \psi)^{\prime
\prime} \right] \not= 0
\]
However, the iterative relation (\ref{nesu}) is suggestive that the negative order densities $\mathcal H_{-n}$ 
determine an independent hierarchy of  conserved
charges that are in involution amongst themselves, and possibly including  $\mathcal H_{1}$ and $\mathcal H_{2}$,
in terms of the Poisson bracket (\ref{ppbrac}).
\begin{equation}
\{ H_{-n} , H_{-m} \}_s \ = \ 0 \ \ \ \ \ \ \ \ \ {\rm for} \ n,m\geq 0
\la{newh}
\end{equation}
We have confirmed this to the densities given in (\ref{hesu}), for example we find
\[
\{ H_{-3}, H_{-4} \}_s = 3 \lambda^2 \int \! ds \, \frac{d}{ds}( \bar \psi \psi ) \ \simeq \ 0
\]
At the moment we lack a simple proof that all the negative order charges are in involution, the integrability
(\ref{newh}) is a conjecture; the fact that the first four charges are in involution strongly suggests
that the conjecture could be correct.

In Figure 1 we summarize schematically the positive and negative order hierarchies, and the correspondence
for the quantities that have a familiar interpretation, together with their Weyl duality relations.

%%%%%%%%%%%%%
%%%%%%%%%%%%%
%%%%%%%%%%%%%%

 \begin{figure}[!hbtp]
  \begin{center}
    \resizebox{7.5cm}{!}{\includegraphics[]{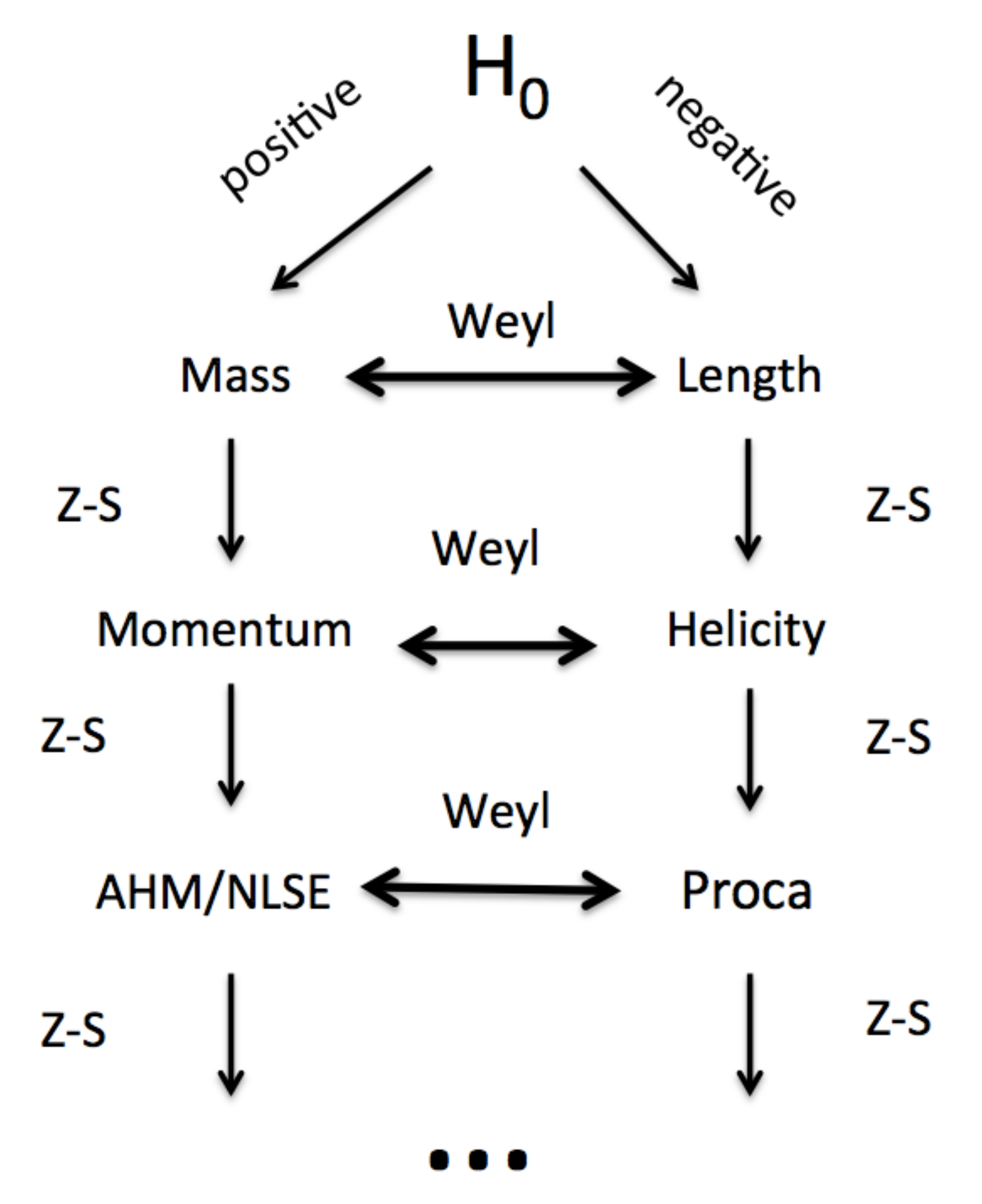}}
    \caption{Schematic relations between those conserved densities that have a natural interpretation
    in terms of gauge field theory,  in the positive order (\ref{hier4}) 
    and negative order (\ref{nesu}), (\ref{hesu}) hierarchies. AHM/NLSE is for Abelian Higgs Model {\it a.k.a.}
    nonlinear Schr\"odinger equation, Weyl denotes relation by the Weyl duality between the hierarchies, 
    and  Z-S stands for Zakharov-Shabat recursive relation. Both hierarchies amalgamate at $ H_0=0$.
     } 
    \label{fig:figure-1}
  \end{center}
\end{figure}

%%%%%%%%%%%%%
%%%%%%%%%%%%%
%%%%%%%%%%%%%%

%%%%%%%%%%%%%%%%%%%%%%%%%%%%%%%%%%%%%%%%%%%%%%%%%%%%%%%%%%
%
%
%
%
%
%
%
%
%
%
%
%
%
%
%
%%%%%%%%%%%%%%%%%%%%%%%%%%%%%%%%%%%%%%%%%%%%%%%%%%%%%%%%%

%%%%%%%%%%%%%%%%%%%%%%%%%%%%%%%%%%%%%%%%%%%%%%%%%%%%%%%%%%
%
%
%
%
%
%
%
%
%
%
%
%
%
%
%
%%%%%%%%%%%%%%%%%%%%%%%%%%%%%%%%%%%%%%%%%%%%%%%%%%%%%%%%%%

\section{Frame  Independence and Gauge Structure }

We  employ the recursion relations (\ref{posu}), (\ref{nesu}) to 
specify Hamiltonian energy functions, that describe  curves and their dynamics 
in $\mathbb R^3$.
The explicit recursion relations that we have presented 
utilize a fixed variable $s\in [0,L]$
that we have chosen to identify with 
the arc-length parameter of the curve. However, we have also
explained how reparametrization covariant generalizations
can be introduced,  
order by order in the recursion. 

Notwithstanding,  in order to describe a curve in $\mathbb R^3$ we need in addition to solve 
the Frenet equation (\ref{DS1}), (\ref{kgn}). The Frenet equation involves
the curvature and torsion, and as such it is frame dependent. 
We proceed to complement reparametrization covariance
with manifest frame independence.

According to (\ref{gautau}), (\ref{gaukappa}) 
the curvature and torsion ($\kappa,\tau$) transform exactly 
like an Abelian Higgs multiplet ($\phi, A$)
under frame rotations, provided  we identify  the Frenet frame curvature
$\kappa$ as the modulus  of
a complex (Higgs) 
scalar field $\phi$, and  the torsion
$\tau$ as the one dimensional U(1) gauge field $A$. In the arc-length parametrization,
the action of frame rotation can thus be summarized as follows:
\[
\begin{array}{ccccc} 
\phi (s) \ & \sim & \ \kappa (s) \ & \to & e^{i\eta(s)} \kappa(s) \\ \\
A(s) \ & \sim & \ \tau(s) \ & \to & \tau(s) -  \eta_s(s) 
\end{array}
\]
Consequently
frame independent Hamiltonians for curves 
can be constructed in parallel with  U(1) gauge invariant Hamiltonians of  the Higgs multiplet
($\phi,A$).  
An example of the latter is the standard Abelian Higgs model Hamiltonian
\begin{equation}
H_3 = \int ds \, \left\{ | ( \frac{d}{ds} + i  A ) \phi |^2 + \lambda|\phi|^4  \right\}
\la{ahm}
\end{equation}
If we identify the Hasimoto variable as follows,
\begin{equation}
\psi (s) = \phi (s)  \exp\{ i \! \int\limits^s \! A \}
\la{psiphi}
\end{equation}
we recover the NLSE Hamiltonian (\ref{nlse}). On the other hand, if in (\ref{ahm})
we introduce the change of variables
\begin{equation}
\begin{array}{lcl}
\kappa \ \simeq \ \phi \  & \to & \ \sigma \, e^ {i \theta } \\ \\
\tau \ \simeq \ A \ & \to & J  = A - \frac{i}{2} \frac{d}{ds} \ln \phi + \frac{i}{2} \frac{d}{ds} \ln \phi^* \\ \\
& & \hskip 0.3cm  \simeq \tau + \frac{d}{ds} \arg \kappa
\end{array}
\la{asu}
\end{equation}
the NLSE Hamiltonian becomes
\begin{equation}
H_3 \ = \ \int ds \left\{ \left(\frac{d \sigma}{ds}  \right)^2 + e^2 \sigma^2 J^2 + \lambda\sigma^4 \right\}
\la{nlse2}
\end{equation}
Like $\psi$ in (\ref{psiphi}), both $J$ and $\sigma$ are manifestly frame (gauge) independent variables; 
$J$ is commonly called the supercurrent variable, in the context of Abelian Higgs model; $\sigma$ is the Higgs
condensate.

The entire NLSE hierarchy can be similarly presented in terms of
the frame independent variables (\ref{asu}). For example, 
for $H_2$ we get from (\ref{h2}) (note: $\sigma = \sqrt{\rho}$)
\begin{equation} 
H_2 \ = \ \int ds \, \sigma^2 J
\la{h22}
\end{equation}
In the Abelian Higgs model, this is the canonical momentum; for $H_1$ we have in (\ref{posu})
\begin{equation}
H_1 \ = \ \int ds \, \sigma^2
\end{equation}
which  is the Higgs mass term, in the Abelian Higgs model.

For the negative order density $\mathcal H_{-2}$  we get from (\ref{hesu})
\[
\mathcal H_{-2}  \ = \ \tau - i \frac{d}{ds} \ln \kappa  \ \simeq \ \tau + \frac{d}{ds} \arg \kappa - i  \frac{d}{ds} \ln \sigma
\]
The   charge is 
\begin{equation}
H_{-2}  \ = \ \int ds \, J
\la{cs}
\end{equation}
We identify (\ref{cs})  as a one dimensional version of helicity, or Abelian Chern-Simons term,  
for the  gauge invariant variable  $J$.
For the density of $H_{-3}$ we get similarly
\[
H_{-3}  =  \left(i \tau -  \frac{d\ln \kappa}{ds}\right)^2   + i \frac{d}{ds} \!\left[  \tau - i \frac{d\ln \kappa}{ds} \right] + 1
\]
\begin{equation}
 \simeq - J^2 + i \frac{d}{ds} J + 1
\la{Hm3} 
\end{equation}
The conserved charge is 
\begin{equation}
H_{-3} \ =  \ \int ds \, J^2   \ + \ L
\la{proca}
\end{equation}
where $L$ is the length of the curve.
In the Abelian Higgs model, we identify the first term in (\ref{Hm3}), (\ref{proca})  
as  the one dimensional version of the gauge invariant {\it i.e.} frame independent
Proca mass, for the variable $J$. 

%%%%%%%%%%%%%%%%%%%%%%%%%%%%%%%%%%%%%%%%%%%%%%%%%%%%%%%%%%
%
%
%
%
%
%
%
%
%
%
%
%
%
%
%
%%%%%%%%%%%%%%%%%%%%%%%%%%%%%%%%%%%%%%%%%%%%%%%%%%%%%%%%%

%%%%%%%%%%%%%%%%%%%%%%%%%%%%%%%%%%%%%%%%%%%%%%%%%%%%%%%%%%
%
%
%
%
%
%
%
%
%
%
%
%
%
%
%
%%%%%%%%%%%%%%%%%%%%%%%%%%%%%%%%%%%%%%%%%%%%%%%%%%%%%%%%%%

\section{Discrete Chains}

In the discrete case of a piecewise linear space polygon {\it i.e.} chain,
the reparametrization {\it i.e.} diffeomorphism  covariance is not of concern.
However, the requirement of invariance under frame rotations persists. 
Consequently, we start
by revisiting the construction of frames for a piecewise linear
polygonal chain in $\mathbb R^3$ \cite{dff}. We take $\mathbf r_i$ ( $i=1,...,N$)
to be the coordinate sites of the vertices, and at each vertex site $i$ we introduce the 
unit tangent vector 
\begin{equation}
\mathbf t_i = \frac{ {\bf r}_{i+1} - {\bf r}_i  }{ |  {\bf r}_{i+1} - {\bf r}_i | }
\la{t}
\end{equation}
We frame the chain by introducing at each vertex $i$ the unit binormal vector
\begin{equation}
\mathbf b_i = \frac{ {\mathbf t}_{i-1} \times {\mathbf t}_i  }{  |  {\mathbf t}_{i-1} \times {\mathbf t}_i  | }
\la{b}
\end{equation}
and the unit normal vector 
\begin{equation}
\mathbf n_i = \mathbf b_i \times \mathbf t_i
\la{n}
\end{equation}
The orthogonal triplet ($\mathbf n_i, \mathbf b_i , \mathbf t_i$) defines the discrete Frenet  
frame at the vertex $\mathbf r_i$ of the chain. 
The  bond angles along the chain are
\begin{equation}
\kappa_{i} \ \equiv \ \kappa_{i+1 , i} \ = \ \arccos \left( {\bf t}_{i+1} \cdot {\bf t}_{i} \right)
\la{bond}
\end{equation}
and the torsion angles are
\begin{equation}
\tau_{i} \ \equiv \ \tau_{i+1,i} \ = \ {\rm sign}\{ \mathbf b_{i-1} \times \mathbf b_{i} \cdot \mathbf t_{i} \}
\cdot \arccos\left(  {\bf b}_{i+1} \cdot {\bf b}_{i} \right) 
\la{tors}
\end{equation}
If these angles are known, we can use  the discrete Frenet equation
\begin{equation}
\left( \begin{matrix} {\bf n}_{i+1} \\  {\bf b }_{i+1} \\ {\bf t}_{i+1} \end{matrix} \right)
= 
\left( \begin{matrix} \cos\kappa \cos \tau & \cos\kappa \sin\tau & -\sin\kappa \\
-\sin\tau & \cos\tau & 0 \\
\sin\kappa \cos\tau & \sin\kappa \sin\tau & \cos\kappa \end{matrix}\right)_{\hskip -0.1cm i+1 , i}
\left( \begin{matrix} {\bf n}_{i} \\  {\bf b }_{i} \\ {\bf t}_{i} \end{matrix} \right) 
\la{DFE2}
\end{equation}
to construct the frame at vertex $i+i$ 
from the frame at vertex $i$. Once we have the frames we get the entire chain, up to global rotations 
and translations, using
\begin{equation}
\mathbf r_k = \sum_{i=0}^{k-1} |\mathbf r_{i+1} - \mathbf r_i | \cdot \mathbf t_i
\la{dffe}
\end{equation}
This amounts to an integration of (\ref{curve}), in the discrete case.
For simplicity, in the sequel we assume that the bond lengths {\it i.e.}
distances between the
neighboring vertices are constant
\begin{equation}
| {\bf r}_{i+1} - {\bf r}_i | = a
\la{a}
\end{equation}
With no loss of generality we can set $\mathbf r_0 = 0$, and choose $\mathbf t_0$ so that it points 
along the positive $z$-axis, so 
\[
\mathbf r_1 = a \, {\hat{\mathbf z}}
\]
In parallel with the continuum curves, (\ref{DFE2}) proposes us to construct energy functions for discrete
chains in terms of the bond and torsion angles.

We note that
(\ref{dffe}) does not involve the vectors $\mathbf n_i$ and $\mathbf b_i$. Thus,  
as in the case of continuous curves, we may introduce
an arbitrary SO(2) rotation between them, without affecting 
the chain. At each vertex site $i$ this frame rotation acts as follows, 
\begin{equation}
 \left( \begin{matrix}
{\bf n} \\ {\bf b} \\ {\bf t} \end{matrix} \right)_{\!i} \!
\rightarrow  \! 
 \left( \begin{matrix}
\cos \Delta_i & \sin \Delta_i & 0 \\
- \sin \Delta_i & \cos \Delta_i & 0 \\ 
0 & 0 & 1  \end{matrix} \right) \left( \begin{matrix}
{\bf n} \\ {\bf b} \\ {\bf t} \end{matrix} \right)_{\! i}
\la{discso2}
\end{equation}
Here the $\Delta_i$ are the arbitrarily chosen local rotation angles. 
On the bond angles,
the effect of frame rotations can be presented as follows,
\begin{equation}
\kappa_i  \, T^2 \ \to \ \kappa_i \, (\cos \Delta_{i+1} T^2 + \sin \Delta_{i+1} T^1 )
\la{kso2}
\end{equation}
This is the real SO(2) version of the complex U(1) rotation 
(\ref{gaukappa}). Thus we may consider  $\kappa_i$ as a complex variable
so that its real part coincides with the $T^2$ component of (\ref{kso2}), and the imaginary
part is the $T^1$ component of (\ref{kso2}). The frame transformation property is
\begin{equation}
\kappa_i \ \to \ e^{i\Delta_{i+1}} \kappa_i
\la{kip2}
\end{equation}
The bond angle (\ref{bond}) is the modulus of this variable. 
On the torsion
angles, the frame rotation has the following effect \cite{dff}
\begin{equation}
\tau_i \ \to \ \tau_i - \Delta_{i+1} + \Delta_{i}
\la{tauig}
\end{equation}

{\it A priori}, the fundamental range of the bond angle $\kappa_i$ is  $ [0,\pi]$. For the 
torsion angle the range is $\tau_i \in [-\pi, \pi)$. Consequently we may 
identify ($\kappa_i, \tau_i$) with the canonical 
latitude and longitude angles of a two-sphere $\mathbb S^2$. 
However, in the sequel we find it useful to extend the range
of $\kappa_i$ into $ [-\pi,\pi]$ $mod(2\pi)$, but with no change in the range of $\tau_i$. 
We compensate for this two-fold covering of $\mathbb S^2$ 
by introducing the following $\mathbb Z_2$ symmetry
\begin{equation}
\begin{matrix}
\ \ \ \ \ \ \ \ \ \kappa_{j} & \to  &  - \ \kappa_{j} \ \ \ \hskip 1.0cm  {\rm for \ \ all} \ \  j \geq i \\
\ \ \ \ \ \ \ \ \ \tau_{i }  & \to &  \hskip -2.5cm \tau_{i} - \pi 
\end{matrix}
\la{dsgau}
\end{equation}
This is a special case of (\ref{discso2}),  with
\[
\begin{matrix} 
\Delta_{l} = \pi \hskip 1.0cm {\rm for} \ \ l \geq i+1 \\
\Delta_{l} = 0 \hskip 1.0cm {\rm for} \ \ l <  i+1 
\end{matrix}
\]

%%%%%%%%%%%%%%%%%%%%%%%%%%%%%%%%%%%%%%%%%%%%%%%%%%%%%%%%%%
%
%
%
%
%
%
%
%
%
%
%
%
%
%
%
%%%%%%%%%%%%%%%%%%%%%%%%%%%%%%%%%%%%%%%%%%%%%%%%%%%%%%%%%

%%%%%%%%%%%%%%%%%%%%%%%%%%%%%%%%%%%%%%%%%%%%%%%%%%%%%%%%%%
%
%
%
%
%
%
%
%
%
%
%
%
%
%
%
%%%%%%%%%%%%%%%%%%%%%%%%%%%%%%%%%%%%%%%%%%%%%%%%%%%%%%%%%%

\section{The discretized nonlinear Schr\"odinger equation}

Two {\it a priori} different integrable discrete versions of  
the continuum NLSE have been presented, the one
in \cite{abla2} and the one in \cite{kore}. We review these models, before we proceed
to derive a discretization of (\ref{recu1c}), (\ref{posu}), using the frame independence of a discrete chain as
our guiding principle.
Following \cite{fadtak}, we denote the two discrete models
by  LNS$_1$ and LNS$_2$, respectively. It can be shown, that LNS$_1$ is equivalent
to the lattice Heisenberg spin chain model (LHM)  \cite{fadtak}. 
In \cite{hoff} it has also been shown, that LNS$_1$ and LNS$_2$ are 
gauge equivalent the sense, that their Lax pair representations can be gauge transformed 
to each other.  But a proof based on gauge equivalence  of the Lax pair  does not give an explicit relation 
between the variables. In particular, it does enable us to reveal, how to describe discrete
curves, transferably  in terms of the two versions of discretized  non-linear Schr\"odinger equation.
 For this reason, we show how to establish 
the equivalence of LHM and LNS$_2$ in terms of an explicit change of variables.

We start by describing the lattice Heisenberg spin chain 
model {\it i.e.} LHM. With $\mathbf s_i $ a three component unit vector
at site $i=1,...,N$, the Hamiltonian is (for notational simplicity
we set bond length $a = 2$ in (\ref{a}) and take the classical spin 
parameter $s$ in \cite{fadtak} to have value $s=1$)
\begin{equation} 
H_{LHM} \ = \ -  \sum\limits_{i=1}^{N-1} \ln ( 1 + \mathbf s_i \cdot \mathbf s_{i+1})
\la{lhm1}
\end{equation}
and the Poisson bracket is the discrete version of (\ref{pbhs}),
\begin{equation}
\left \{ s_i^a , s_j^b \right\} = - \epsilon^{abc} s_j^c \delta_{ij}
\la{dpb2}
\end{equation}
The equation of motion  for $\mathbf s_i$ is the Landau-Lifschitz equation
\begin{equation}
\frac{ d \mathbf s_i}{dt} = \{ H_{LHM} , \mathbf s_i \} = - \mathbf s_i \times \frac{ \partial H_{LHM}}{\partial \mathbf s_i}
\la{llm}
\end{equation}
It reproduces (\ref{hc2}) in the continuum limit $a\to 0$.

We may use (\ref{lhm1}) to introduce an
energy function for discrete polygonal chains. For this we simply
identify the spin variable $\mathbf s_i$ with the tangent vector $\mathbf t_i$, defined  in (\ref{t}). Note that since only
$\mathbf s_i \sim \mathbf t_i$ appears in (\ref{lhm1}), (\ref{dpb2})  
the energy function (\ref{lhm1})  determines chain dynamics which is 
manifestly invariant under the local frame rotations (\ref{discso2}). 
In terms of the discrete Frenet frame,
the equation of motion (\ref{llm}) becomes
\begin{equation}
\frac{ d \mathbf t_i}{dt} = \frac{1}{2} \left( \tan \frac{\kappa_{i+1}}{2} \, 
\mathbf b_{i+1} - \tan \frac{\kappa_i}{2} \, \mathbf b_i \right)
\la{LLE}
\end{equation}
so that for our polygonal chain we obtain, from (\ref{t}),  the discrete version of the localized induction approximation
(\ref{lia2}),
\begin{equation}
\frac{d\mathbf r_i}{dt} = \, \tan \frac{\kappa_i}{2} \, \mathbf b_i
\la{lia}
\end{equation}
Since the Landau-Lifschitz equation is integrable there are conserved quantities, equal in number with the 
canonical coordinates (or momenta). 
In particular, we note that the equation of motion (\ref{LLE}) preserves the closure condition of the chain
\[
\frac{d}{dt}\left\{  \sum\limits_{i=1}^{N} \mathbf t_i \right\}  \ = \ 0
\]
This is the discrete version of the  continuum statement that $H_{-1}$ in (\ref{L}), (\ref{nesu}) is a conserved quantity.

In terms of the Frenet frame
bond (\ref{bond}) and torsion (\ref{tors}) 
angles, with  ($\kappa_{i+1,i}, \tau_{i+1,i}$) $\equiv$ ($\kappa_{i}, \tau_{i}$) the Poisson bracket (\ref{dpb2}) is
\begin{equation}
\begin{matrix}
\{ \kappa_i , \kappa_{i+1} \} \hskip 0.3cm & = & \sin \tau_{i+1} \\ \\
\{ \kappa_{i-2} , \tau_i \} & = & -\cos \tau_{i-1} \csc \kappa_{i-1} \\ \\
\{ \kappa_{i-1} , \tau_i \} & = & \cot \frac{\kappa_{i-1}}{2} \, + \, \cos \tau_i \cot \kappa_i \\ \\
\{ \kappa_i , \tau_i \}  & = & - \cot \frac{\kappa_i}{2} - \cos \tau_i \cot \kappa_{i-1} \\ \\
\{ \kappa_{i+1}, \tau_i \} & = & \cos \tau_{i+1} \csc \kappa_i \\ \\
\{ \tau_{i-1} , \tau_i \} & = & \csc \kappa_{i-1} \, ( \sin \tau_i \cot \kappa_i + \sin \tau_{i-1} \cot \kappa_{i-2} ) \\ \\
\{ \tau_{i-1}, \tau_{i+1} \} & =  &  \sin \tau_i \csc \kappa_{i-1} \csc \kappa_i
\end{matrix}
\la{dpb3}
\end{equation}
The Hamiltonian is
\begin{equation}
H \ = \ - \sum\limits_i \ln ( 1 + {\mathbf t}_i \cdot {\mathbf t}_{i+1} ) \ = \ -2 \sum\limits_{i} \ln \cos \frac{\kappa_i}{2}
\la{Hln}
\end{equation}
and the equation of motion (\ref{LLE}) is
\begin{equation}
\frac{d\kappa_i}{dt} = \tan \frac{\kappa_{i-1}}{2} \sin \tau_i - \tan\frac{\kappa_{i+1}}{2} \sin \tau_{i+1}
\la{LLE1}
\end{equation}
\[
\frac{d\tau_i}{dt} = \cos\tau_i \, \left( \cot \kappa_i \tan \frac{\kappa_{i-1}}{2} - \cot \kappa_{i-1} \tan \frac{\kappa_i}{2} \right)
\]
\begin{equation}
+ \tan \frac{\kappa_{i+1}}{2} \csc \kappa_i  \cos \tau_{i+1} - \tan\frac{\kappa_{i-2}}{2} \csc \kappa_{i-1} \cos\tau_{i-1}   
\la{LLE2}
\end{equation}
From  (\ref{LLE2}) we immediately compute
\[
\frac{d}{dt} \sum\limits_i \tau_i = 0
\]
This is the discrete version of the conservation of total torsion $H_{-2}$ in
(\ref{T}). We also note that  in the continuum limit, (\ref{Hln})
becomes (\ref{K}).

We now show that  the equations of motion (\ref{LLE1}), (\ref{LLE2}) 
coincide with those of the LNS$_2$ model. For this
we introduce the following discrete variable,
\begin{equation}
\psi_i  = \tan \frac{\kappa_i}{2} \, e^{i \vartheta_i}
\la{hmd1}
\end{equation}
where we have the (anti)symmetrized combination
\[
\vartheta_i = \frac{1}{2} \left( \sum\limits_{k=1}^i \tau_k - \sum_{k=i+1}^N \tau_k \right)
\]
Note that (\ref{hmd1}) does not remain invariant under (\ref{kip2}), (\ref{tauig}). 
It should not be interpreted as the proper discretized version 
of the frame independent Hasimoto variable (\ref{psisym}).
Consequently we proceed by explicitly assuming the Frenet framing.

A direct computation, using (\ref{dpb3}), 
gives
\[
\frac{d\psi_i}{dt} = \frac{1}{2} \sec^2 \frac{\kappa_i}{2} e^{i\vartheta_i} 
\frac{d\kappa_i}{dt} + i \tan \frac{\kappa_i}{2} e^{i\vartheta_i}
\frac{d\vartheta_i}{dt} 
\]
On the other hand,
\[
(1+| \psi |^2) (\psi_{i+1} + \psi_{i-1} )
\ 
= \ \sec^2 \frac{\kappa_i}{2} \left( \tan \frac{\kappa_{i+1}}{2} e^{ i \tau_{i+1}} +
\tan \frac{\kappa_{i-1}}{2} \, e^{-i\tau_i} \right) \, e^{ i\vartheta_i }
\]
By absorbing a factor of 2 in the definition of $t$ and redefining 
\[
\psi_i \ \to \ \psi_i e^{i t}
\]
we obtain the LNS$_2$ equation \cite{abla2} 
\[
i \frac{ d\psi_i}{dt} = - \left( \psi_{i+1} - 2 \psi_i + \psi_{i-1} \right) - |\psi_i|^2 ( \psi_{i+1} + \psi_{i-1} )
\]
This is a Hamiltonian equation of motion, with 
\[
H = -\sum\limits_i \left( \psi_i \psi^\star_{i+1} + \psi^\star_i \psi_{i+1} \right)  
\ = \ -2 \sum\limits_i \tan\frac{\kappa_i}{2}
\tan \frac{\kappa_{i+1}}{2} \cos \tau_{i+1}
\]
and Poisson brackets 
\[
\left\{ \psi_i , \psi^\star_j \right\} = i \left( 1+|\psi_i|^2 \right) \delta_{ij} 
\]
\[
\left\{ \psi_i , \psi_j \right\} = 
\left\{ \psi_i^\star ,  \psi^\star_j \right\} = 0
\]
These relations define the LNS$_2$ model. Thus we have established 
the equivalence between the LHM and LNS$_2$ models, by a direct computation
in terms of the discrete Frenet frames.

%%%%%%%%%%%%%%%%%%%%%%%%%%%%%%%%%%%%%%%%%%%%%%%%%%%%%%%%%%
%
%
%
%
%
%
%
%
%
%
%
%
%
%
%
%%%%%%%%%%%%%%%%%%%%%%%%%%%%%%%%%%%%%%%%%%%%%%%%%%%%%%%%%

%%%%%%%%%%%%%%%%%%%%%%%%%%%%%%%%%%%%%%%%%%%%%%%%%%%%%%%%%%
%
%
%
%
%
%
%
%
%
%
%
%
%
%
%
%%%%%%%%%%%%%%%%%%%%%%%%%%%%%%%%%%%%%%%%%%%%%%%%%%%%%%%%%%

\section{Discretized Recursion Relations Of The Positive Order}

As an energy function for a discrete chain, the  LHM (\ref{lhm1}), (\ref{dpb2}) engages only the tangent vector
$\mathbf t_i \simeq \mathbf s_i$ and consequently it is, as such, manifestly frame independent. But the combined variable
(\ref{hmd1}) that relates LHM and LNS$_2$ does not transform covariantly under frame rotations.  Consequently
it can not be interpreted as a discrete frame invariant version of the Hasimoto variable.
Our goal, in the following, is to derive a proper {\it i.e.} manifestly frame independent  
discretized version of the continuum NLSE hierarchy (\ref{recu1c}) of conserved charges.  These can be utilized as
energy functions, in the case of piecewise linear polygonal chains. 

As the guiding principle we shall utilize {\it solely} 
the concept of invariance under frame rotations, in terms of the
formal identification as gauge transformations: 
We demand that the discretized versions of the conserved charges 
should be manifestly 
invariant under the local frame rotations
(\ref{discso2}) {\it i.e.} they should derive from a gauge principle.
They should go over to their continuum NLSE hierarchy
counterparts, including the perfect derivative contributions, 
and smoothly in the limit where the bond length $a$ 
between the neighboring vertices (\ref{a}) vanishes. 

We note that the {\it primary emphasis} here  is on ensuring
frame independence  rather than on preserving integrability. The former is a necessary
requirement for constructing meaningful energy functions for discrete chains,
while the latter is a bonus.   We presume that if the discretized recursion relations
are carefully constructed so that frame invariance is preserved, these relations
will also lead to integrability, if indeed the latter can be consolidated with the former. 
For this, we shall confirm that in a carefully executed 
continuum limit we indeed obtain
the conserved charges of the NLSE hierarchy, including the perfect derivative contributions.
In combination with frame independence, this is a strong indication
that the discrete hierarchies we present are integrable.

We introduce the following discrete version of the
Hasimoto variable (\ref{psi}),
\begin{equation}
\psi _{i}=\kappa _{i}\exp\{ ia\sum_{k=1}^{i}\tau _{k+1,k}\}  
\la{hasi}
\end{equation}%
%We interpret (\ref{hasi}) so that it has support at the $i^{th}$ vertex.
Note that (\ref{hasi}) is different from (\ref{hmd1}). The latter is not manifestly frame independent, while (\ref{hasi})
is constructed to be so.
We continue to identify the $\kappa_i \equiv \kappa_{i+1,i}$ as the bond angles (\ref{bond}).
But from the  conceptual point of view, there is now an advantage to consider
the bond angles 
$\kappa_i$ as variables that have support at the  corresponding $i^{th}$ vertex. 

Similarly, the variables $\tau_{k+1,k}$ are related to the chain geometry, as torsion angles
according to (\ref{tors}). Unlike the bond angles that reside at the vertices, the $\tau_{k+1,k}$  are now 
interpreted as variables that have 
support on the link that connects the vertices $k$ and $k+1$.

As geometric bond and torsion angles,
($\kappa_i, \tau_{i+1,i}$) relate to the discrete Frenet framing of the chain. Again, in the general frames,
we extend $\kappa_i$ into a complex variable.  The site dependent frame rotation 
that sends
\begin{equation}
\begin{array}{lcl}
\kappa_{i+1,i} \ \equiv \ \kappa _{i} & \rightarrow & \kappa _{i}\exp\{i\Delta_{i+1}\} \\ \\
~~~~~ \tau_{i+1,i} & \rightarrow & \tau _{i+1,i}- \frac{1}{a} (\Delta _{i+1}-\Delta_{i})
\end{array}
\la{u1-transformation}
\end{equation}
leaves the Hasimoto variable (\ref{hasi}) intact (with $\Delta_1 =0$). Note that in relation to
(\ref{tauig}), we have  re-defined the torsion angles by scaling them with the  bond length $a$.

We shall now introduce the discrete analogs of the densities $\omega_n(s)$ in (\ref{recu1}). 
We denote them $w_{n}[i]$.  We propose to evaluate the $w_{n}[i]$  from the 
following discretization of the (general frame) recursion relation (\ref{recu1c}),
\begin{equation}
w_{n+1}[i]= \frac{i}{a}\left\{\, w_{n}[i]-e^{-ia\tau _{i+1,i}}w_{n}[i-1]\, \right\}
\ + \ \lambda \bar\kappa _{i}\sum_{l<n}w_{l}[i]w_{n-l}[i],
\la{Qn}
\end{equation}
The recursion starts with
\begin{equation}
\begin{array}{lcl}
w_{0}[i] & = & 0 \\ \\
w_{1}[i] & = & \frac{1}{4} \kappa_{i} \ \simeq \ \frac{1}{4} \kappa_{i+1,i}
\end{array}
\la{initQ}
\end{equation}
We have here carefully constructed (\ref{Qn}) using the standard lattice gauge theory \cite{gernot} discretization of covariant
derivative. We note that there are variants of this discretization \cite{gernot}. In particular we may introduce
\[
w_{n+1}[i]= \frac{i}{a}\left\{\, w_{n}[i+1]e^{ia\tau _{i+2,i+1}}- w_{n}[i]\, \right\}
+ \lambda \bar\kappa _{i}\sum_{l<n}w_{l}[i]w_{n-l}[i],
\]
But for the present purposes it is sufficient to consider the version (\ref{Qn}).
We note that with the initialization (\ref{initQ}), the quantities $w_n[i]$ transform
according to
\[ 
w_n[i] \ \to \ e^{i \Delta_{i+1}} w_n[i]
\]
under discrete frame rotations. In particular, (\ref{Qn}) is covariant under frame rotations.

In analogy with (\ref{Hnp}), we introduce the following frame invariant quantities,
\begin{equation}
N_{n}\left[ i\right] =  \bar\kappa _{i}w_{n}\left[ i\right]
\la{conserved-quantities}
\end{equation}%
and in the following we proceed by using
the Frenet framing where $\bar\kappa_i \equiv \kappa_i$.
We verify that in the continuum limit $a\to 0$ the ensuing  quantities (\ref{conserved-quantities}) 
reproduce the  densities (\ref{posu}), including the perfect derivative contributions. In combination with discrete
frame rotation covariance, this will support
the  conjecture, that (\ref{conserved-quantities})
are the proper discrete versions of the continuum conserved NLSE charges.
We confirm this explicitly,  for the four densities in (\ref{hier4}).

In the case of  
$w_{1}[i]$, it is clear that in the
continuum limit  the quantity $N_{1}\left[ i\right] $ is reduced to%
\begin{equation}
\lim_{a\rightarrow 0}N_{1}\left[ i\right] =\frac{\kappa ^{2}}{4},  \label{N1}
\end{equation}%
which coincides with the density $\mathcal H_1$ in (\ref{hier4}). 

For $w_{2}\left[ i\right] $, we get
\begin{eqnarray}
w_{2}\left[ i\right]  &=&\frac{i}{a}\left[ w_{1}[i]-e^{-ia\tau
_{i+1,i}}w_{1}[i-1]\right]   \notag \\
&=&\frac{i}{a}\left[ \frac{\kappa _{i}}{4}-e^{-ia\tau _{i+1,i}}\frac{\kappa
_{i-1}}{4}\right]   \la{Q2}
\end{eqnarray}%
In the continuum limit, this gives
\begin{equation}
\lim_{a\rightarrow 0}N_{2}\left[ i\right] = \frac{\tau \kappa ^{2}}{4} - \frac{i}{8}
\frac{d}{ds} \left( \kappa^2\right)
 \la{N2}
\end{equation}%
which is the conserved density $\mathcal H_2(s)$ of NLSE hierarchy, including the perfect derivative contribution.

The $w_{3}\left[ i\right] $ reads%
\[
w_{3}[i] =\frac{i}{a}\left\{\, w_{2}[i]-e^{-ia\tau _{i+1,i}}w_{2}[i-1]\, \right\}
+ \lambda\kappa _{i}w_{1}[i]w_{1}[i]  
\]
\[
= \frac{i}{a}\left\{ \frac{i}{a}\left[ \frac{\kappa _{i}}{4}-e^{-ia\tau
_{i+1,i}}\frac{\kappa _{i-1}}{4}\right] \right.
\]
\begin{equation}
\left. - e^{-ia\tau _{i+1,i}}\left( \frac{i}{a%
}\left[ \frac{\kappa _{i-1}}{4}  -e^{-ia\tau _{i,i-1}}\frac{\kappa _{i-2}}{4}%
\right] \right) \right\} + \lambda \kappa _{i}\frac{\kappa_{i}}{4}\frac{\kappa _{i}}{%
4}
\la{Q3}
\end{equation}%
In the continuum limit, 
%\[
%\lim_{a\rightarrow 0}Q_{3}[i] = 
%\lim_{a\rightarrow 0}\frac{1}{a}\left\{ \frac{1}{a}\left[ \frac{\kappa
%_{i}}{4}-\frac{2\kappa _{i-1}}{4}+ \frac{\kappa _{i-2}}{4}\right]
%\right.
%\]
%\[
%- \left[
%i\tau _{i,i-1}\frac{\kappa _{i-1}}{2}-i\tau _{i-1,i-2}\frac{\kappa _{i-2}}{4}%
%-i\tau _{i,i-1}\frac{\kappa _{i-2}}{4}\right] 
%\]
%\[
%\left.  - a\tau _{i,i-1}\tau _{i-1,i-2}%
%\frac{\kappa _{i-2}}{4}\right\} + i \lambda \frac{\kappa ^{3}}{16} 
%\]
%\begin{equation}
%= \frac{\kappa ^{\prime \prime }}{4}-i\frac{\tau \kappa ^{\prime }}{2}-i%
%\frac{\tau ^{\prime }\kappa }{4}-\frac{\tau ^{2}\kappa }{4}+i \lambda \frac{\kappa ^{3}%
%}{16}  \la{Q3-continuum}
%\end{equation}%
\[
\lim_{a\rightarrow 0}N_{3}\left[ i\right]  = - \frac{1}{4} \kappa \kappa^{\prime
\prime } - \frac{i}{2} \tau \kappa^{\prime }\kappa  - \frac{i}{4} \tau^{\prime}\kappa^{2} + \frac{1}{4} \tau^{2}\kappa^{2} + 
\frac{\lambda}{16}\kappa^{4} 
\]
\begin{equation}
= \left( \frac{\kappa \kappa ^{\prime }}{4}-i\frac{\tau \kappa ^{2}}{4}%
\right) ^{\prime } + \frac{\left( \kappa ^{\prime }\right) ^{2}}{4}+ \frac{\tau
^{2}\kappa ^{2}}{4}+\frac{\lambda}{16}\kappa^{4}
\la{N3}
\end{equation}%
This coincides with the density of the nonlinear Schr\"{o}dinger Hamiltonian, together with the
perfect derivative term in (\ref{hier4}).

For the fourth quantity,%
\begin{equation}
w_{4}[i]=\frac{i}{a}\left\{\, w_{3}[i]-e^{-ia\tau _{i+1,i}}w_{3}[i-1]\, \right\}
+2\kappa _{i}w_{2}[i]w_{1}[i] 
\la{Q4}
\end{equation}
%Using the previously derived results we obtain  
%\[
%\lim_{a\rightarrow 0}Q_{4}[i]=-i\frac{3}{16}\kappa ^{3}\tau +\frac{5}{16}%
%\kappa ^{2}\kappa ^{\prime }+i\frac{\kappa \tau ^{3}}{4} - \frac{3\kappa \tau
%\tau ^{\prime }}{4} 
%\]
%\[
%- i\frac{\kappa \tau ^{\prime \prime }}{4} -\frac{3\kappa
%^{\prime }\tau ^{2}}{4}-i\frac{3\kappa ^{\prime }\tau ^{\prime }}{4}-i\frac{%
%3\tau \kappa ^{\prime \prime }}{4}+\frac{\kappa ^{\prime \prime \prime }}{4}
%\]%
In the continuum limit, we have verified that the ensuing charge $N_4[i]$ reproduces $\mathcal H_4(s)$ in 
(\ref{hier4}), inclusive the perfect derivative term.

%%%%%%%%%%%%%%%%%%%%%%%%%%%%%%%%%%%%%%%%%%%%%%%%%%%%%%%%%%
%
%
%
%
%
%
%
%
%
%
%
%
%
%
%
%%%%%%%%%%%%%%%%%%%%%%%%%%%%%%%%%%%%%%%%%%%%%%%%%%%%%%%%%

%%%%%%%%%%%%%%%%%%%%%%%%%%%%%%%%%%%%%%%%%%%%%%%%%%%%%%%%%%
%
%
%
%
%
%
%
%
%
%
%
%
%
%
%
%%%%%%%%%%%%%%%%%%%%%%%%%%%%%%%%%%%%%%%%%%%%%%%%%%%%%%%%%%

\section{Discretized Recursion Relations Of The Negative Order}

We proceed to the discrete version of the negative order hierarchy (\ref{nesu}).
We propose the following  manifestly frame rotation invariant discretization,
\begin{equation}
w_{-(n+1)}\left[ i\right] =-\frac{i}{a\kappa^2_i}\left\{ \,  \kappa^2_i w_{-n}\left[ i\right] 
\ -\kappa^2_{i-1}e^{-ia\tau
_{i+1,i}}w_{-n}\left[ i-1\right] \, \right\} 
%\]
%\begin{equation}
+\lambda\kappa^3
_{i}\sum_{l<n}w_{-l}[i]w_{-(n-l)}[i]  \label{Q-negative}
\end{equation}%
As before, 
\[
w_{0}\left[ i\right] = 0
\] 
and we initialize the recursion by setting
\[
w_{-1}\left[ i\right] = \frac{1}{\kappa _{i}} \label{Q1-negative}
\]
This ensures that $w_{-(n+1})[i]$ transforms covarianty under discrete frame rotations,
\[
w_{-n}[i] \ \to \ e^{-i\Delta_{i+1}} w_{-n}[i]
\]
The higher order negative order quantities  $N_{-n}\left[ i\right] $ are defined as%
\begin{equation}
N_{-n}\left[ i\right] = \kappa _{i}w_{-n}\left[ i\right]  
\la{N-negative}
\end{equation}
In the following we proceed using the Frenet frames; by construction, the final
results are frame independent.

We have explicitly confirmed, that in the continuum limit this produces  the correct Frenet frame 
densities in (\ref{hesu}),
 including the perfect derivative contributions. For example, we immediately confirm that
\begin{equation*}
N_{-1}=1,
\end{equation*}%
Similarly, for $w_{-2}\left[ i\right] $ we have%
\begin{equation}
w_{-2}[i] =-\frac{i}{a \kappa_i^2}\left\{ \kappa _{i}^2 w_{-1}[i] - \kappa^2 _{i-1} 
e^{-ia\tau_{i+1,i}} w_{-1}[i-1]\right\}
 \la{Q2-negative}
\end{equation}%
In the continuum limit, (\ref{N-negative}) gives immediately 
\begin{equation}
\lim_{a\rightarrow 0}N_{-2}\left[ i\right] = \tau - i (\ln \kappa )^\prime  
\la{Q2-negative-continuum}
\end{equation}%
which coincides with the density $\mathcal H_{-2}$ in (\ref{hesu}), including the perfect derivative.
Similarly, we find that  $N_{-3}[i]$ produces the Proca mass term $\mathcal H_{-3}(s)$ in (\ref{hesu}) including
the perfect derivative term: We have
\[
\omega_{-3}[i] = - \frac{i}{2 \kappa_i^2} \left( \kappa_i^2 w_{-2}[i] - \kappa_{i-1}^2 e^{-i a \tau_i} w_{-2}[i-1]\right)
\ + \ \lambda \kappa_i^3 w_{-1} [i] w_{-1}[i]
\]
We have also confirmed by explicit computation, that 
$N_{-4}[i]$ produces $\mathcal H_{-4}(s)$ in (\ref{hesu}), including the perfect derivative. 
Consequently, we conjecture that this is the case to all negative order terms.
 
Finally, we note that  if we introduce the discrete version of the Weyl transformation
\[
N_{-n}[i] \ \to e^{\theta[i]} N_{-n} [i] \ \equiv \  N^\theta_{-n} [i] 
\]
with
\[
\theta[i] = \ln |\kappa_i|^2
\] 
we find that in the continuum limit we recover the densities with positive order,
\[
N_{-n}[i] \to N^\theta_{-n} [i] \ \simeq N_n[i] \ \buildrel{a \to 0}\over{\longrightarrow} \mathcal H_n(s)
\]
We have confirmed this explicitly, to order $n=4$.

%%%%%%%%%%%%%%%%%%%%%%%%%%%%%%%%%%%%%%%%%%%%%%%%%%%%%%%%%%
%
%
%
%
%
%
%
%
%
%
%
%
%
%
%
%%%%%%%%%%%%%%%%%%%%%%%%%%%%%%%%%%%%%%%%%%%%%%%%%%%%%%%%%

%%%%%%%%%%%%%%%%%%%%%%%%%%%%%%%%%%%%%%%%%%%%%%%%%%%%%%%%%%
%
%
%
%
%
%
%
%
%
%
%
%
%
%
%
%%%%%%%%%%%%%%%%%%%%%%%%%%%%%%%%%%%%%%%%%%%%%%%%%%%%%%%%%%

\section{Energy functions for discrete chains}

In \cite{oma1}-\cite{oma5}, the following 
energy function has been introduced and applied 
to model folded  poteins as space filling piecewise linear polygonal chains,
\begin{equation}
E = \sum\limits_{i=1}^N \! \left\{ - 2 \sigma_{i+1} \sigma_i + b \, \sigma_i^2 J_i^2 + c( \sigma_i^2 - m^2)^2 \right\}
\ + \ \sum\limits_{i=1}^N \! \left\{ d \, J_i + q\, \sigma_i^2 J_i + e\, J_i^2 \right\} \ \ \ \ \ \ (\sigma_{N+1} = 0)
\la{ulfen}
\end{equation}
The functional form of this energy function is justified by a
{\it naive} discretization of the following linear 
combination of charges $H_a \ (a=-3,-2,1,2,3)$, that appear in the continuum 
NLSE hierarchy and its Weyl dual hierarchy; see equations (\ref{nlse2})-(\ref{proca})
\begin{equation}
E =  \int \! ds \left\{ \left(\frac{d \sigma}{ds}  \right)^2 + e^2 \sigma^2 J^2 + \lambda\sigma^4 + \mu^2 \sigma^2 \right\}
\ + \ \int\! ds \left\{ d \, J + q \sigma^2 J + e \, J^2 \right\}
\la{Econt}
\end{equation}
%As a U(1) gauge theory, this corresponds to the following extension of (\ref{ahm}),
%\[
%E = \int\limits_0^L \! ds \, \left\{ | ( \frac{d}{ds} + i  A ) \phi |^2 + \lambda|\phi|^4 +  
%\right\}
Here the frame independent {\it i.e.} 
gauge invariant supercurrent variables are utilized. The energy function (\ref{Econt}) has a natural gauge theory
interpretation: The first line is the Abelian Higgs model in the supercurrent variables, 
with spontaneously broken potential term when $\mu^2<0$. 
The second line displays the Chern-Simons term, 
the canonical momentum and the Proca mass term, respectively.
When a {\it naive} discretization of the derivative contribution is implemented,
\[
\left(\frac{d\sigma}{ds}\right)^2  \ \to \ (\sigma_{i+i}-\sigma_i)^2 \ \sim \ 2\sigma_i^2 - 2 \sigma_{i+1} \sigma_i
\]
we clearly arrive at (\ref{ulfen}). 
This discretization of the energy function (\ref{Econt}) appears to correctly describe various properties
of folded proteins \cite{oma2}-\cite{oma5}.

Unfortunately, (\ref{ulfen}) also leads to a conceptual predicament. As the bond angle, the variable $\sigma_i$ 
in (\ref{ulfen}) is by construction a non-negative quantity, and in particular the first term in (\ref{ulfen}) does not remain 
invariant under a 
local transformation that send $\sigma_k \to - \sigma_k$ at one given site. On the other hand,
(\ref{Econt})  is even in
$\sigma$ and consequently in this continuum 
energy function, $\sigma(s)$ can be extended into a real valued variable.
In the discrete case, we have previously used the $\mathbb Z_2$
symmetry transformation (\ref{dsgau}) to extend the bond angle variable $\sigma_i \sim \kappa_i$ 
from positive values into both positive and negative values. This symmetry transformation leaves 
the vectors $\mathbf t_i$ intact, and as a consequence the discrete chain
constructed from (\ref{dffe}) is also  intact. Consequently, from the point of view of discrete chains,
it might be desirable  to have a discrete energy function that remains  
invariant under this symmetry transformation. Unfortunately, (\ref{ulfen}) does not have this property.

We note that the continuum version of the  $\mathbb Z_2$ transformation
corresponds to a discontinuous frame flipping by $\pi$  at an inflection point of the curve; the continuum
version of (\ref{dsgau}) engages both the unit step and the Dirac $\delta$-function \cite{dff}.  

Our goal is to {\it minimally} improve (\ref{ulfen}) into a discrete energy function which is constructed from
the densities in the discrete NLSE hierarchy, so that the ensuing energy 
has the local $\mathbb Z_2$ symmetry  (\ref{dsgau}). In parallel with (\ref{Econt}), we are instructed to try and proceed 
by using the discretized versions of the charges $H_a \ (a=-3,-2,1,2,3)$. These charges have been derived 
in the previous two Sections, these are the quantities $N_a[i]; \ \ a \in \mathbb Z$. These quantities 
are by construction
fully frame independent and consequently invariant under  the local symmetry transformation  (\ref{dsgau}). 
Consequently we start by introducing the $\mathbb Z_2$ invariant discrete combination
\begin{equation}
\alpha_3 N_3[i] + \alpha_2 N_2[i] + \alpha_1 N_1[i] + \alpha_{-2} N_{-2}[i] + \alpha_{-3} N_{-3}[i]
\la{Ei}
\end{equation}
By construction, this overlays the putative energy function.
In general the charges $N_a[i]$ have both a real part and an imaginary part,
\[
N_a[i] = Re \, N_a[i] + i Im \, N_a[i] 
\]
and these are both independently invariant under the $\mathbb Z_2$ transformation; 
we have observed (\ref{hier4}), (\ref{hesu}) that in the continuum limit, the imaginary parts are
perfect derivatives, except for a term in $\mathcal H_{-3}$.  
As a consequence, whenever  $N_a[i]$ is complex, 
we can interpret both the real part and the imaginary part, separately, as
contributions to the energy. Accordingly, we account for them independently by substituting 
\begin{equation}
N_a[i]  \to \  \beta_1\, Re \,N_a[i] \ + \  \beta_2 \, Im\, N_a[i] \ \ \ \ \ \ \beta_{1,2} \in \mathbb R
\la{NRC}
\end{equation}
in the expansion (\ref{Ei}). The ensuing real-valued quantity can be used as an energy function:
According to the results in the previous Sections, its continuum limit contains
the density in (\ref{Econt}), in addition of terms that are perfect derivatives. 
In this way we can construct a discrete energy function that extends  (\ref{ulfen}) in a 
manner which ensures invariance under the $\mathbb Z_2$ transformation (\ref{dsgau}). 

Unfortunately, (\ref{Ei}) is quite elaborate in comparison with (\ref{ulfen}).
But we have found that a discrete energy density which extends (\ref{ulfen}) in a manner which is
invariant under (\ref{dsgau}) and yields the density in (\ref{Econt}) in the continuum, can also be obtained by 
considering only the three quantities $N_a[i]$ with $a=-2,1,2$.

We start with $N_1[i]$. It gives the  following mass term contribution, we now work 
in the general frame where the $\kappa_i$ are complex
\begin{equation}
E_1[i] \ = \  m^2 \bar \kappa_i \kappa_i 
\la{E1}
\end{equation}
with $m$  a parameter.

We proceed to $N_2[i] $ in (\ref{Q2}). 
We obtain the contribution 
\begin{equation}
N_2[i] \sim \bar\kappa_i \kappa_i  - \bar \kappa_{i}e^{-i a \tau_{i+1,i} } \kappa_{i-1}
\la{N2}
\end{equation}
The first term can be combined with (\ref{E1}), it amounts to a redefinition of  the {\it a priori} free
parameter  $m$.  

Following Section VII, we conceptually
identify the second term as the standard U(1) gauge invariant kinetic term 
that is commonly introduced  in lattice Abelian Higgs model \cite{gernot}. For this,  we simply interpret
\[
\kappa_i \sim \phi_{i+1}
\]
as the complex lattice 
Higgs field, and 
\[
\tau_{i+1,i} \sim A_{i+1,i}
\] 
the ensuing gauge field so that  the second term is the familiar \cite{gernot}
\[
\bar \phi_{i+1} e^{ia A_{i+1,i}} \phi_i
\]
We write the second term in  (\ref{N2}) as follows,
\[
 \bar \kappa_{i}e^{-i a \tau_{i+1,i} } \kappa_{i-1} \ = \ |\kappa_i|^2 \exp\{ -i a \tau_{i+1,i} + \ln \frac{\kappa_{i-1}}{\kappa_i} \}
\]
\begin{equation}
= \ \sigma_i \sigma_{i-1} \exp\{- i a J_{i+1,i}\}
\la{Ji}
\end{equation}
where, in analogy with (\ref{asu}), we have introduced the gauge {\it i.e.} frame independent variables
\[
\begin{array}{ccc} 
\sigma_i & = &   \hskip -4.2cm | \kappa_i |  \\ \\
J_{i+1,i} & = & \tau_{i+1,i} - \frac{1}{a} \{ \arg(\kappa_{i-1}) - \arg(\kappa_i) \}
\end{array}
\]
The quantity (\ref{Ji})  has both a real part
and an imaginary part, that are independently frame rotation invariant.  Consequently, in line with (\ref{NRC}), we obtain two
independent quantities that we utilize to introduce the following contribution to discrete energy density
\begin{equation}
E_{2}[i] =  \sum\limits_i \{ -2  \sigma_i \sigma_{i-1} \cos ( a J_{i+1,i} ) +  \frac{q}{a} \sigma_i \sigma_{i-1} \sin ( a J_{i+1,i})\}
\la{E2ab}
\end{equation}
where the parameters are chosen for easy comparison with (\ref{ulfen}).

Following Section VIII we proceed and 
extend the variables $\sigma_i$ from their {\it a priori} positive to arbitrary real values. For this
we  introduce the local $\mathbb Z_2$ symmetry (\ref{dsgau}), that acts on the present variables 
at a given site $k$ as follows: 
\begin{equation}
\begin{matrix}
\ \ \ \ \ \ \ \ \ \sigma_{j} & \to  &  - \ \sigma_{j} \ \ \ \hskip 1.0cm  {\rm for \ \ all} \ \  j \geq k \\
\ \ \ \ \ \ \ \ \ \ J_{k+1,k }  & \to &  \hskip -2.5cm J _{k+1,k} - \frac{\pi}{a} 
\end{matrix}
\la{dsgau2}
\end{equation}
Unlike (\ref{ulfen}), the energy (\ref{E2ab}) remains intact under this symmetry, thus we can utilize it to
extend  $\sigma_i$ in (\ref{E2ab}) from positive values to arbitrary real values.

When we expand (\ref{E2ab}) in the bond length parameter $a$, we get
\[
\approx -2 \sigma_{i}\sigma_{i-1} +  q \sigma_{i}\sigma_{i-1} J_{i+1,i} + a^2 \sigma_i \sigma_{i-1} J^2_{i+1,i} 
+ \mathcal O(a^3)
\]
which is reminiscent of the first, fifth and second terms in (\ref{ulfen}), respectively. The difference is a
higher order correction in the bond length parameter $a$.

We now proceed to the negative order term $N_{-2}[i]$. From  (\ref{Q2-negative}) we 
extract the following discrete energy density contribution. As in (\ref{NRC}) we
account for  the real and imaginary part separately, and write
\begin{equation}
E_{-2}[i] = -\frac{2e}{a^2} \frac{\sigma_i}{\sigma_{i-1} }\cos ( a J_{i+1,i} ) + \frac{d}{a} \frac{\sigma_i}{\sigma_{i-1} }
\sin ( a J_{i+1,i})
\la{E2c}
\end{equation}
\[
= \frac{d}{a} \sin ( a J_{i+1,i}) -\frac{2e}{a^2} \cos ( a J_{i+1,i} )  
\]
\begin{equation}
+ \frac{1}{\sigma_{i-1}} \left( \frac{ \sigma_i - \sigma_{i-1}}{a}\right)
\left\{ d \sin ( a J_{i+1,i}) - \frac{2e}{a} \cos ( a J_{i+1,i} )\right\}
\la{E2c2}
\end{equation}
Again, this remains intact under the $\mathbb Z_2$ transformation (\ref{dsgau2}) so that we are able to take
$\sigma_i$ to be a real valued quantity. When we expand  in the bond length parameter $a$, the first line gives
modulo a constant
\[
\approx  d J_{i+1,i} + e J^2_{i+1,i} + \mathcal O(a^2)
\]
which is reminiscent of  the fourth and sixth terms in (\ref{ulfen}) with corrections that are of higher order in $a$.
The second line in (\ref{E2c2}) gives
\[
- \frac{2e}{a}\, \frac{1}{\sigma_{i-1}} \left( \frac{ \sigma_i - \sigma_{i-1}}{a}\right)
\ + \  \left( \frac{ \sigma_i - \sigma_{i-1}}{\sigma_{i-1}}\right)\left( d J_{i+1,i} + e J_{i+1,i}^2 \right) + \mathcal O(a^2) 
\]
where, in the continuum limit, the first term becomes a perfect derivative,
\[
\frac{1}{\sigma_{i-1}} \frac{ \sigma_i - \sigma_{i-1}}{a} \ \to \ \frac{d\ln \sigma}{ds} 
\]
and the second term vanishes in this limit, as $\mathcal O(a)$

Finally, the quartic self-interaction of $\sigma_i$ arises naturally from (\ref{Ei}), see (\ref{Q3}), (\ref{N3}).
Here, we propose to add it essentially {\it ad hoc}, as the square of the density (\ref{E1}); any even polynomial
of $\kappa_i$ is invariant under the discrete transformation (\ref{dsgau2}) and as such admissible in energy.
Most likely this is in violation of integrability. But in applications to proteins, 
it is desirable to have soliton solutions for which (discrete) translation invariance 
becomes broken, by a Peiers-Nabarro barrier \cite{naba}.

We combine (\ref{E1}), (\ref{E2ab}) and (\ref{E2c}) to arrive at the following 
improved and $\mathbb Z_2$ invariant version of the energy density (\ref{ulfen}),
\[
E \ = \ \sum_{i=1}^N \! \left\{ - 2 \sigma_i \sigma_{i-1} \cos(a J_{i+1,i} ) + \frac{q}{a} \sigma_i \sigma_{i-1}
\sin(a J_{i+1,i}) \right\}
\ + \ \sum_{i=1}^N \! c ( \sigma_i^2 - m^2 )^2 
\]
\begin{equation}
+ \ \sum_{i=1}^N \! \left\{  \frac{d}{a} \, \frac{\sigma_i}{\sigma_{i-1}} \sin(a J_{i+1,i} ) - 
\frac{2e}{a^2} \frac{\sigma_i}{\sigma_{i-1}} \cos(a J_{i+1,i} ) \right\}
\la{efin}
\end{equation}
It would be interesting to see, how (\ref{efin}) models folded proteins.

%%%%%%%%%%%%%%%%%%%%%%%%%%%%%%%%%%%%%%%%%%%%%%%%%%%%%%%%%%
%
%
%
%
%
%
%
%
%
%
%
%
%
%
%
%%%%%%%%%%%%%%%%%%%%%%%%%%%%%%%%%%%%%%%%%%%%%%%%%%%%%%%%%

%%%%%%%%%%%%%%%%%%%%%%%%%%%%%%%%%%%%%%%%%%%%%%%%%%%%%%%%%%
%
%
%
%
%
%
%
%
%
%
%
%
%
%
%
%%%%%%%%%%%%%%%%%%%%%%%%%%%%%%%%%%%%%%%%%%%%%%%%%%%%%%%%%%

\section{Summary}

In summary, we have addressed the problem how to utilize extrinsic geometry, to 
derive Hamiltonian energy functions for continuous and discrete 
strings that move in three dimensional space. The approach we have chosen, is based on a simple geometric
principle: We insist, that the energy density must remain invariant under local frame rotations and, in the
case of continuous strings,  transform covariantly under reparametrizations. In this manner
we arrive naturally to energy densities, that relate to the integrable hierarchy of the nonlinear Schr\"odinger
equation. Furthermore, we have found that there is a Weyl dual to this hierarchy, that also 
relates to the energy densities that are relevant for strings in three space dimensions. 
We have argued that this  additional hierarchy is also integrable, and we have confirmed
the integrability explicitly in the case of  the first few nontrivial quantities. 
In addition, we have established, by explicit computation, the
equivalence of the two known integrable discrete nonlinear Schr\"odinger hierarchies. We have also shown, how a 
discretized generalization of the nonlinear Schr\"odinger equation that has been previously shown to describe
folded proteins, can be obtained as a particular limit of 
the quantities that appear in the discretized Zakharov-Shabat recursion relations that we have proposed.

We hope that our approach can provide a systematic basis for the general 
description of both continuous and discrete string-like configurations in three 
space dimensions, in particular space filling ones.  

\section{Acknowledgements:}

A.J.N. thanks L. Faddeev  and N. Molkenthin for discussions. S.H. thanks Dai Jin for discussions.
This work has been supported by CNRS PEPS grant, Region Centre 
Rech\-erche d$^{\prime}$Initiative Academique grant,
Sino-French Cai Yuanpei Exchange Program (Partenariat Hubert Curien), and 
Qian Ren Grant at BIT. Y.J. also  acknowledges the financial support from 
the NSFC under Grant No. 11275119.

\end{document}